\documentclass{scrartcl}

\usepackage{amsmath}	% Für Matheumgebungen wie $ $,...
\usepackage{amssymb}	% Mathematische Symbole
\usepackage{a4wide}	

\usepackage{graphicx}	% Für Bilder
\usepackage{float}	% Für Gleitumgebungen (Bilder mit Bildunterschriften)
\usepackage{cite}

\newcommand{\chisq}{$\chi^2$~}

\begin{document} 
\title{Delineating parton distributions and the strong coupling}
\author{P.~Jimenez-Delgado$^a$ and E.~Reya$^b$}
\date{\large{$^a$ Thomas Jefferson National Accelerator Facility\\
Newport News, VA 23606, USA\\
\vspace{0.3cm}
$^b$Institut f\"ur Physik, Technische Universit\"at Dortmund\\
D-44221 Dortmund, Germany}}

\maketitle
\vspace{-10cm}
{\begin{flushright}
DO-TH 14/03 \\
JLAB-THY-14-1853\\
Phys. Rev.~{\bf D89}, 074049 (2014)
 \end{flushright}}
\vspace{+10cm}
\thispagestyle{empty}

{\begin{center}{\bf{Abstract}}\end{center}}
\begin{abstract}
Global fits for precision determinations of parton distributions, together
with the highly correlated strong coupling $\alpha_s$, are presented 
up to next-to-next-to-leading order (NNLO) of QCD utilizing most world
data (charm and jet production data are used where theoretically
possible), except TeVatron gauge boson production data and LHC data
which are left for genuine {\em predictions}.
This is done within the `dynamical' (valencelike input at $Q_0^2=0.8$ GeV$^2$)
and `standard' (input at $Q_0^2=2$ GeV$^2$) approach.  
The stability and reliability of the results are ensured by including nonperturbative
higher-twist terms, nuclear corrections as well as target mass corrections,
and by applying various ($Q^2,W^2$) cuts on available data.  
In addition, the $Q_0^2$ dependence of the results is studied in detail.
Predictions are given, in particular for LHC, on gauge- and Higgs-boson as 
well as for top-quark pair production.  
At NNLO the dynamical approach results in $\alpha_s(M_Z^2)=0.1136\pm0.0004$,
whereas the somewhat less constrained standard fit gives $\alpha_s(M_Z^2)=0.1162\pm0.0006$. 
\end{abstract}

\newpage

\section{Introduction}

Precision determinations of parton distribution functions (PDFs), 
together with the strongly correlated strong coupling $\alpha_s$,
are of fundamental importance for testing the expectations and 
predictions of the standard model (SM) of the strong, electromagnetic
and weak interactions, such as the production of weak gauge bosons $(W^{\pm},\, Z^0)$, of
heavy quarks $h\bar{h}\, (h=c,b,t)$, of hadronic jets and,
in particular, of the Higgs-boson at hadron colliders (TeVatron, LHC). A 
high statistical accuracy of PDFs is equally mandatory for
exploring and delineating the limits of the SM, thus establishing 
possible signatures of approaches which go beyond the SM.
\bigskip

First we shall concentrate on the extraction of PDFs from presently 
available data within the {\em{standard}} approach,
followed by most groups up to next-to-next-to-leading order (NNLO) of 
QCD \cite{ref1,ref2,ref3,ref4,ref5,ref6,ref7,ref8,ref9}, 
where the input scale is fixed
at {\em{one}} arbitrarily chosen value $Q_0^2\geq 1$ GeV$^2$.
Besides the perturbative leading-twist $\tau = 2$ framework (for recent 
reviews see, for example, \cite{ref10,ref11,ref12}) with
its strongly correlated $\alpha_s(Q^2)$, nonperturbative higher-twist 
(HT) $\tau=4$ and possible $\tau=6$ contributions are
included as well, giving rise to power-like suppressed corrections 
proportional to $(Q^2)^{-(\tau-2)/2}$.
Furthermore, target-mass and nuclear deuteron corrections are also taken 
into account. In order to learn about the stability, and
thus the reliability of the results, we perform fits to various subsets 
of data by applying various kinematic $(Q^2,W^2)$ cuts
on the available data where $W^2=Q^2(\frac{1}{x}-1)+M^2$.
In particular, the required HT contributions turn out to be sensitive to 
the chosen $Q^2$ and $W^2$ (or $x$) range of the data
fitted, and therefore the requirement of stable HT terms in size and 
shape plays a non-negligible role in all analyses.
In addition we study the dependence of the results on the specific 
choice of the input scale $Q_0^2$ (ranging from 1 to 9 GeV$^2$
in current analyses), which in most cases has not been systematically 
addressed so far.
In principle the results should not depend on these choices; in 
practice, however, a dependence develops as a consequence of what
is being referred to as `procedural bias' \cite{ref13}.
We shall start from our previous standard leading twist-2 next-to-leading order (NLO) GJR 
\cite{ref14} and NNLO JR09 \cite{ref2} analyses in order to
learn about the stability and reliability of these PDFs and the values 
of $\alpha_s$ obtained.
\bigskip

Next, we turn alternatively to the {\em{dynamical}} PDFs which are 
generated from input distributions at an optimally determined
low input scale $Q_0^2 < 1$ GeV$^2$ (\cite{ref2,ref14}
and references therein).
This has the advantage that the input distributions {\em{naturally}} tend to 
{\em{valencelike}} functions, i.e., not only the valence but also
the sea and gluon input densities {\em{vanish}} at small $x$, due to the 
conventional renormalization group (RG) evolutions.
Thus the behavior of parton distributions and deep inelastic scattering 
(DIS) structure functions at $Q^2 > Q_0^2$ and at small $x$
is entirely {\em{predictable}} as a consequence of QCD dynamics (gluon 
and quark radiation), while in the {\em{standard}}
approach it has to be fitted.
Indeed, the very characteristic and unique {\em{steep}} small-$x$ 
{\em{pre}}dictions at $x \lesssim 10^{-2}$ were subsequently
confirmed, about 20 years ago, by HERA DIS experiments 
($\!\!$\cite{ref2,ref14} and references therein).
These dynamical PDFs, as well as the corresponding $\alpha_s(Q^2)$, will 
be reanalyzed along very similar lines as discussed
above for obtaining PDFs within the standard approach.
It should be noted that the input gluon distribution for $Q_0^2$ smaller than
about 1 GeV$^2$ turns out to be always {\em valencelike}. 
In contrast, however, present precision data do not favor a valencelike sea
distribution as well:  we refrain from enforcing a valencelike sea since for
$Q_0^2<0.8$ GeV$^2$ the fitted value of $\chi^2$ strongly increases 
and $\alpha_s(M_Z^2)$ decreases substantially, despite taking into account
HT contributions.
Therefore, in order not to deteriorate the quality of the fit and to avoid any 
bias towards low $\alpha_s$ values, we shall fix the dynamical input scale to 
be $Q_0^2=0.8$ GeV$^2$ where the sea increases slightly at very small $x$.
This solution will nevertheless be referred to as dynamical, since it is the 
valencelike (input) gluon distribution which dominates the small-$x$ behavior
of structure functions.
\bigskip

Section 2 is devoted to a presentation and discussion of various theoretical issues, 
such as the target-mass corrections to structure functions, the formulation of 
higher-twist contributions, nuclear corrections and  the evaluation of the 
statistical uncertainties of PDFs.
In Section 3 we present our quantitative {\em{standard}} and 
{\em{dynamical}} results for structure functions, PDFs and hadronic jet production.
Our calculations basically refer to the so-called `fixed-flavor number scheme 
(FFNS) where heavy quarks ($c,b,t$) are not considered
as massless partons within the nucleon, i.e.~the number of active 
(light) flavors $n_f$ appearing in the splitting functions
and corresponding Wilson coefficients will be fixed, $n_f=3$.
This scheme is fully predictive in the heavy quark sector (i.e., without 
any additional model assumptions) where the heavy
quark flavors are produced entirely perturbatively from the initial 
light (u,d,s) quarks and gluons with the full heavy quark mass
$m_{c,b,t}$ dependence taken into account in the production cross 
sections - as required experimentally, in particular, in
the threshold region.
In many situations, calculations within this factorization scheme become 
unduly complicated (for a recent discussion see,
for example, \cite{ref3,ref15}), and in many cases even 
impossible due to the unknown fully massive ($m_{c,b,t}\neq 0$)
matrix elements at NNLO and even NLO.
For this reason we shall generate, from our unique FFNS 
PDFs, effective PDFs in the so-called
`variable-flavor number scheme' (VFNS) where the heavy quarks ($c,b,t$) are 
considered to be massless partons within the nucleon
as well.
Here, the required NLO and NNLO (massless) cross sections are available in the 
literature for a variety of important production processes,
thus allowing for {\em{pre}}dictions for gauge- and SM Higgs-boson production 
experiments at TeVatron and ongoing LHC measurements.
It should be emphasized that, on purpose, we do not include TeVatron 
gauge-boson production data and LHC data in our
fitting procedure which therefore allows for genuine {\em{pre}}dictions 
of these measurements in Section 4.
Thus we can explicitly test the reliability and usefulness of the QCD 
improved parton model.
Finally, our conclusions are summarized in Section 5.

%%%%%%%%%%%%%%%%%%%%%%%%%%%%%%%%%%%%%%%%%%%%%%%%%%%%%%%%%%%%%%%

\section{Theoretical Issues}
A description of the underlying QCD framework (FFNS, VFNS) used in our (JR) analyses has already been given
and summarized in \cite{ref2,ref3} up to NNLO and will not be repeated here. 
However, since the last determination of our PDFs there have been some experimental and theoretical developments which deserve consideration and partly motivate the present investigations. 
\bigskip

One of the changes in our approach is that for the SLAC \cite{ref16}, NMC \cite{ref17} and BCDMS \cite{ref18,ref19} data we use now the directly measured cross sections, instead of the extracted structure functions and ratios of structure functions, since problems with such extractions have been encountered \cite{ref20} (note that we have already done this for the HERA data). Additional assumptions on the spread of the beam energy would be required for reconstructing the measured cross section from the extracted E665 structure functions \cite{ref21}; we have kept these data since a systematic error for the published structure functions has been provided which takes into account possible inaccuracies in the extraction (assumed value of $R=\sigma_L/\sigma_T$).
\bigskip

This implies that, in addition to the well-known nucleon target-mass corrections (TMCs) for $F_2$ \cite{ref22} already included in our previous analyses, we include now TMCs for $F_L$ as well,  which have been calculated \cite{ref10,ref23} in the same (operator product expansion) framework:
%equation 1 and 2
\begin{align}
F_2^{\rm TMC}(n,Q^2) & \equiv \int_0^1dx\, x^{n-2}F_2^{\rm TMC}(x,Q^2)\nonumber\\
& = \sum_{j=0}^3 \left(\frac{M^2}{Q^2}\right)^j\, \frac{(n+j)!}{j!(n-2)!}\,\,\frac{F_2(n+2j,Q^2)}{(n+2j)(n+2j-1)} + {\cal{O}}\left(\left(\frac{M^2}{Q^2}\right)^4\right)
\end{align}
\begin{align}
F_L^{\rm TMC}(n,Q^2) & \equiv \int_0^1 dx\, x^{n-2}F_L^{\rm TMC}(x,Q^2)\nonumber\\
& = \sum_{j=0}^3 \left(\frac{M^2}{Q^2}\right)^j \, \frac{(n+j)!}{j!n!}\left[F_L(n+2j,\, Q^2)+\frac{4j}{(n+2j)(n+2j-1)}\, F_2(n+2j,\, Q^2)\right]\nonumber\\
&  +{\cal{O}}\left(\left(\frac{M^2}{Q^2}\right)^4\right)
\end{align}
with the longitudinal structure function in Bjorken-$x$ space being given by
$F_L(x,Q^2) = (1+4x^2M^2/Q^2)F_2-2xF_1$.
\bigskip

Another issue raised by the ABM collaboration \cite{ref5} is the necessity of including higher-twist contributions in the description of fixed-target data, even if moderate kinematic cuts are used to select the data included in the fits, in particular for the SLAC and NMC data \cite{ref16,ref17}.
The kinematic cuts in our GJR and JR analyses \cite{ref2,ref14} were $Q^2 \geq 4$ GeV$^2$ in virtuality and $W^2\geq 10$ GeV$^2$ in DIS invariant mass squared, and were applied to the $F_2$ values extracted from different beam energies and combined. 
The description was good for the NMC data \cite{ref17} and rather poor for SLAC data \cite{ref16}. 
However, since the number of data points of these experiments was rather small (about 100 data points for NMC and 50 for SLAC), the values of the cuts did not affect much the results in \cite{ref2} or\cite{ref14}. 
This picture changes if data on the cross sections for individual energies are used, which amounts to hundreds of data points for each experiment. 
Higher twist (HT) contributions are introduced in the present investigations through the following phenomenological parametrizations:
%equation 3
\begin{equation}
F_i^{{\rm{HT}},N}(x,Q^2) = F_i^{{\rm TMC}, N}(x,Q^2) + \frac{h_i^N(x)}{Q^2} + \frac{h^{\prime N}_i(x)}{Q^4}
\label{HT}
\end{equation}
where $i=2,L$ and $N=p,n$. 
The function $h$ (respectively $h^\prime$) is a cubic spline which interpolates between a set $\{x_j,h(x_j)\}$ of points, with $x_j$ chosen to be $0,\, 0.1,\, 0.3,\, 0.7,\, 1$ for $F_2$ and $0,\, 0.2,\, 0.6,\, 1$ for $F_L$.
This choice provides enough flexibility for the additional fit parameters $h(x_j)$ with respect to the data analyzed.  
Different choices yield similar results.
Furthermore we put  $h(1)=0$ due to kinematic constraints, and $h(0)=0$ since power-like terms appear
to be not required by HERA data at small $x$.
A possible scale dependence of the $h$ and $h^\prime$ functions has been neglected.
\bigskip

As mentioned before, instead of ratios of structure functions we include now a wealth of deuteron cross section data \cite{ref16,ref17,ref19,ref21}, so that an appropriate description of the nuclear structure is of more relevance than in our previous analyses.
We use nuclear corrections provided by the CJ group \cite{ref24} with the deuteron structure functions given by
%equation 4
\begin{equation}
F_i^d(x,Q^2) = \sum_{N=p,n} \int_{y_{min}}^{y_{max}} dy\, f_{N/d}(y,\gamma)\, F_i^N(\tfrac{x}{y},Q^2) +\;\; \delta^{\rm off} F_i^d(x,Q^2) +\;\; \delta^{\rm shad} F_i^d(x,Q^2)
\end{equation}
where again $i=2,L$ and $\gamma^2 = 1 + 4x^2 \tfrac{M^2}{Q^2}$.
The ``smearing functions'' $f_{N/d}$ are computed from the deuteron wave functions based on particular nucleon-nucleon potentials (we use the Paris potential \cite{ref25}), and implement nuclear binding and Fermi motion effects \cite{ref26,ref27}.
The additional additive terms represent off-shell \cite{ref26} and nuclear shadowing \cite{ref28} corrections (these have been considered only for the $i=2$ case). For the E605 \cite{ref29} Drell-Yan data, as well as for the NuTeV \cite{ref30} and CCFR \cite{ref31} data on dimuon production in $\nu$N DIS, the nuclear corrections of nDS \cite{ref32} have been used. These nuclear corrections were obtained using previous sets of (3-flavor) parton distributions which are consistent with our present determinations, and therefore are especially suited for our analysis.

\bigskip

Yet another improvement in the present analyses is the use of the running-mass definition for DIS charm and bottom production which results in an improved stability of the perturbative series \cite{ref33}. At NLO the heavy quark coefficient functions are exactly known \cite{ref34,ref35}.
Beyond NLO the relevant coefficients are not known exactly and it has become customary to use an approximation obtained via threshold
resummation and exact asymptotic ($Q^2\gg m_h^2$) three-loop results as given in \cite{ref33,ref36,ref37}.
In order to avoid any possible ambiguities, we will rather not use DIS heavy quark data for our nominal NNLO fits.  We shall come
back to this point and to the relevance of such approximations in Sec.~3.
In contrast to our previous analyses \cite{ref2,ref3,ref14}, the factorization scale for heavy quark production will, for consistency, be chosen to be $\mu_F^2=Q^2+4m_h^2$ for the relevant $h=c,b$ flavors. For all our fits we use fixed values of $m_c(m_c) =  1.3 \;{\rm GeV}$, and $m_b(m_b) =  4.2 \;{\rm GeV}$, and furthermore $m_t =  173 \;{\rm GeV}$.
\bigskip

As discussed already in Sec.~1, calculations within the FFNS become unduly complicated in many
situations and even impossible due to the unknown fully massive ($m_h\neq 0$) sub-cross sections.
For this reason we generate, starting from our unique $n_f=3$ flavor FFNS PDFs, efffective PDFs
in the VFNS up to $n_f$=5 where the `heavy' quark distributions $xc$ and $xb$ become massless
PDFs of the nucleon as well (for a recent discussion up to NNLO see, for example, \cite{ref3}).
We use these results for calculating jet as well as gauge- and Higgs-boson production rates.
\bigskip

In addition to the above improvements in the theoretical computations, a novelty in our current analysis, as compared to the published ones in \cite{ref2,ref3,ref14}, is a complete treatment of the systematic uncertainties of the data including experimental correlations. 
The estimator of least squares that we use to take them into account has been explicitly written down in Appendix B of \cite{ref38}; we repeat it here for the convenience of the reader, and to fix the notation for further discussions.
The global \chisq\!\! function is obtained as the sum of the functions of each of the data sets included in the analysis.
These consist in general of $i=1,\ldots,N$ data points of central value $D_i$, total uncorrelated error $\Delta_i$ (statistical and uncorrelated systematic errors added in quadrature), and correlated systematic errors $\Delta_{ij}$ for $j=1,\ldots,M$ sources. 
Denoting the respective theoretical predictions by $T_i$, the \chisq function for a data set is
%equation 5
\begin{equation}
 \chi^2 = \sum_{i=1}^{N} \frac{1}{\Delta_i^2}\left( D_i + \sum_{j=1}^{M} r_j \Delta_{ji} - T_i \right)^2 + \sum_{j=1}^M r_j^2
\label{chi2def}
\end{equation}
where the optimal \emph{systematic shifts} $r_j$ are not additional free parameters but rather are determined analytically as \cite{ref38}
%equation 6
\begin{equation}
r_j = - \sum_{k=1}^{M} A_{jk}^{-1} B_k\, ,\hspace{1em} B_j = \sum_{i=1}^N\Delta_{ji}\frac{D_i-T_i}{\Delta_i^2} \, ,\hspace{1em} A_{jk} = \delta_{jk} + \sum_{i=1}^N \frac{\Delta_{ji}\Delta_{ki}}{\Delta_i^2}
\label{shifts}
\end{equation}
where $\delta_{jk}$ denotes here the Kronecker delta. 
Note that in this expression the errors have been regarded as independent quantities.
This is usually true for the statistical (counting) errors, while the systematic errors are often proportional to the central values, which requires a careful treatment. It is well known that naively scaling those errors as proportional to the measured central values leads to a biased estimation of the theory (towards smaller theoretical predictions), while merely taking them as proportional to the calculated central values in 
Eq.(\ref{chi2def}) leads to biased estimations in the opposite direction (towards larger theoretical predictions). 
An optimal fit can be achieved by scaling the multiplicative systematic errors as proportional to a \emph{fixed} (initial) theory, and, proceeding iteratively, this theory can be self-consistently chosen to be the outcome of the estimation itself (cf.~\cite{ref13} for more details).
Note that this method is used for the combination of inclusive HERA data as well \cite{ref39}.
%%%%%%%%%%%%%%%%%%%%%%%%%%%%%%%%%%%%%%%%%%%%%%%%%%%%%%%%%%%%%%%

\section{Quantitative Standard and Dynamical Results}
From the experimental side, besides results coming from the LHC which will not be considered for the moment, the combined (H1 + ZEUS) HERA data on neutral current and charged current DIS inclusive cross-sections \cite{ref39} and charm electroproduction \cite{ref40} have been published and supersede previous sets. We include as well the model-independent (based on measurements at different energies) data on $F_2$ and $F_L$ from H1 \cite{ref41}, and from recently analyzed BCDMS data as well as from JLab-E99118 and SLAC-E140x measurements \cite{ref42}. As stated in the previous section, we have extended our theoretical description and included higher-twist contributions to the DIS structure functions, which allow us to include Jefferson Lab cross-section measurements \cite{ref43} at relatively low $Q^2$ and $W^2$ values, in addition to the other fixed-target DIS measurements \cite{ref16,ref17,ref18,ref19,ref21,ref44} already included in our previous analyses.
\bigskip

The dynamical determination of strange parton distributions has been investigated up to NLO in \cite{ref45}. There the most precise data on dimuon production from NuTeV \cite{ref30} were found in good agreement with the predictions derived from our GJR08 \cite{ref14} dynamical parton distributions, which have been generated entirely radiatively starting from vanishing strange input distributions. The relevant calculation of the inclusive DIS cross section for charged current charm production is not available at NNLO; however by using the NLO expressions with NNLO distributions the would-be NNLO predictions based on vanishing strange input distributions seem to undershoot the data by more than one would expect for the size of the NNLO corrections. Thus, for lack of a better alternative, we include now the dimuon data of NuTeV \cite{ref30} and CCFR \cite{ref31}, and appropriately extend our parametrizations to include strange-quark distributions at the input scale. Note that this also allows for the
determination of a (non-perturbative) asymmetry in the strange input distributions (a perturbative asymmetry is generated by the RG evolution beyond NLO); see \cite{ref45} and references therein for more details.
\bigskip

The nominal set of data used in our global analyses is further composed of Drell-Yan dilepton production from E605 \cite{ref29}, as well as proton-proton and proton-deuteron \cite{ref46} and ratios of proton-deuteron to proton-proton rates from E866 \cite{ref47}. Furthermore, as in our previous NLO analyses \cite{ref14}, we include (updated) TeVatron jet data \cite{ref48,ref49} in our current NLO analyses; in addition we include now DIS inclusive jet production data from HERA \cite{ref50,ref51}. Since the full NNLO calculations for jet cross sections are still not available, we do not include jet data in our nominal NNLO fits.
(In the purely gluonic channel to dijet production the NNLO/NLO \mbox{K-factor} turns out to be approximately flat across the $p_T$ range corresponding to about a 20\% increase compared to the NLO cross section \cite{ref52}.)
However, in order to illustrate the effects of the incorrect inclusion of those data beyond NLO we carry out fits in which the data are included and the theoretical description is based on NNLO parton distributions with NLO (improved NLO in the case of TeVatron) matrix elements; these fits will be denoted by NNLO*. 
It should be mentioned that such an illustration might be very deceptive since the uncertainties due to the (incorrect) order of PDFs
are small in comparison to the theoretical uncertainty on (the incorrect order of) matrix elements \cite{ref53}.
A complete list of data sets considered in our current analyses is given in Table 1. 
Note that the number of data points and $\chi^2$ values quoted there refer to the nominal cuts imposed on DIS data, for which we require $Q^2 \geq 2 \;{\rm GeV}^2$ and $W^2 \geq 3.5 \;{\rm GeV}^2$. No kinematic cuts were applied to Drell-Yan or jet production data, although some outliers with large errors at the limit of the kinematic coverage have been excluded from the data in \cite{ref46,ref48,ref49}.
\bigskip

\begin{table}[t!]
\centering
\begin{tabular}{|l|c|c c c|c c c|}
\multicolumn{2}{c}{}&\multicolumn{3}{c}{$Q_0^2 = 0.8 \;{\rm GeV}^2$}&\multicolumn{3}{c}{$Q_0^2 = 2 \;{\rm GeV}^2$}\\
\hline
                               & NDP & NLO  & NNLO & NNLO*&  NLO & NNLO & NNLO* \\
\hline
HERA $\sigma$\cite{ref39}      & 621 & 1.44 & 1.26 & 1.38 & 1.29 & 1.19 & 1.25 \\
H1 $F_2$ \cite{ref41}          & 63  & 1.02 & 0.96 & 0.94 & 0.83 & 0.88 & 0.86 \\
H1 $F_L$ \cite{ref41}          & 63  & 1.15 & 1.14 & 1.08 & 1.71 & 1.38 & 1.25 \\
HERA $\sigma^c$ \cite{ref40}   & 52  & 1.41 &  --  & 1.68 & 1.79 &  --  & 2.17 \\
\hline
SLAC p \cite{ref16}            & 504 & 1.20 & 1.16 & 1.12 & 1.14 & 1.11 & 1.10 \\
SLAC d \cite{ref16}            & 517 & 1.13 & 1.10 & 1.07 & 1.08 & 1.07 & 1.06 \\
BCDMS p \cite{ref18}           & 351 & 1.17 & 1.17 & 1.18 & 1.18 & 1.19 & 1.19 \\
BCDMS d \cite{ref19}           & 254 & 1.14 & 1.16 & 1.13 & 1.15 & 1.15 & 1.13 \\
NMC p, d \cite{ref17}          & 516 & 1.52 & 1.53 & 1.55 & 1.51 & 1.52 & 1.53 \\
NMC d/p \cite{ref44}           & 177 & 0.87 & 0.86 & 0.86 & 0.86 & 0.85 & 0.85 \\
E665 p, d \cite{ref21}         & 106 & 1.28 & 1.37 & 1.37 & 1.28 & 1.39 & 1.39 \\
JLab p \cite{ref43}            & 91  & 1.19 & 1.19 & 1.19 & 1.19 & 1.19 & 1.19 \\
JLab d \cite{ref43}            & 91  & 1.19 & 1.24 & 1.24 & 1.22 & 1.25 & 1.25 \\
\hline
BCDMS $F_2$ \cite{ref42}       & 10  & 1.32 & 1.33 & 1.19 & 1.11 & 1.10 & 1.01 \\
BCDMS $F_L$ \cite{ref42}       & 10  & 0.43 & 0.42 & 0.40 & 0.40 & 0.39 & 0.39 \\
SLAC 140x $F_2$ \cite{ref42}   &  2  & 0.39 & 0.19 & 0.28 & 0.83 & 0.44 & 0.53 \\
SLAC 140x $F_L$ \cite{ref42}   &  2  & 1.21 & 1.59 & 1.58 & 1.26 & 1.70 & 1.63 \\
\hline
dimuon \cite{ref30,ref31}      & 180 & 0.57 & 0.56 & 0.56 & 0.56 & 0.55 & 0.55 \\
\hline
E605   \cite{ref29}            & 136 & 1.11 & 1.06 & 1.05 & 1.08 & 1.04 & 1.03 \\
E866 pp \cite{ref46}           & 138 & 1.13 & 1.10 & 1.10 & 1.13 & 1.11 & 1.11 \\
E866 pd \cite{ref46}           & 159 & 1.71 & 1.65 & 1.60 & 1.71 & 1.62 & 1.59 \\
E866 pd/pp \cite{ref47}        & 39  & 1.00 & 1.05 & 1.02 & 1.04 & 1.04 & 1.03 \\
\hline
CDF jet \cite{ref48}           & 64  & 2.72 &  --  & 1.91 & 1.97 &  --  & 1.61 \\
D0 jet \cite{ref49}            & 96  & 1.41 &  --  & 1.14 & 1.09 &  --  & 1.03 \\
ZEUS jet \cite{ref50}          & 30  & 0.67 &  --  & 0.76 & 0.61 &  --  & 0.66 \\
H1 jet \cite{ref51}            & 24  & 1.68 &  --  & 1.59 & 1.97 &  --  & 1.77 \\
\hline
total         & 4296 \// 4030  & 1.25 & 1.19 & 1.21 & 1.22 & 1.17 & 1.19 \\
\hline
\end{tabular}
\caption{Data sets, number of used points, and $\chi^2$ values (per data point) obtained in some of the different variants of our dynamical ($Q_0^2 = 0.8 \;{\rm GeV}^2$) and standard ($Q_0^2 = 2 \;{\rm GeV}^2$) analyses. As far as the total values of $\chi^2$ are concerned it should be noted that the data sets for our NLO and NNLO fits differ slightly.  For illustration we also display the NNLO* results which
correspond to NNLO fits using (incorrectly) NLO matrix elements (sub-cross sections) for observables (charm and jet production) where
NNLO QCD corrections are not fully known yet.}
\label{data}
\end{table}

We have investigated the stability of the results under variations in these cuts. Although the effects in the parton distributions themselves are more moderate, the repercussion on the determination of the higher-twist contributions and of $\alpha_s(M_Z^2)$ is quite significant. To start with, we have considered the possibility of including twist-6 contributions to the DIS structure functions, $h^\prime$ in Eq.(\ref{HT}). However present DIS data do not well constrain these functions: setting them free produces abnormally small values of $\chi^2$ and $\alpha_s(M_Z^2)$ in conjunction with large compensating twist-4 and twist-6 contributions (this was also observed in \cite{ref54}), presumably due to overfitting and blurring the scaling violations. A similar situation occurs if one uses larger cuts (like $Q^2 \geq 4 \;{\rm GeV}^2$, $W^2 \geq 10 \;{\rm GeV}^2$ and $Q^2,W^2 \geq 7 \;{\rm GeV}^2$) and attempts to determine simultaneously the twist-4 contributions and the parton distributions (twist-2).
On the contrary, if the same larger cuts are used but only twist-2 contributions are considered, $\alpha_s(M_Z^2)$ tends to larger values \cite{ref5} of about 0.117 to 0.118 in our NNLO fits, and $\chi^2$ increases considerably; thus these kinds of tighter cuts should preferably be avoided. Fortunately, stable results under smaller variations in the cuts are achieved if one either uses twist-2 contributions only together with rather stringent cuts ($Q^2,W^2 \gtrsim 10 \;{\rm GeV}^2$), or \mbox{twist-2} and twist-4 contributions with lower cuts similar to our nominal choice. Since the data at lower values of $Q^2$ and $W^2$ provide valuable information, in particular in the large-$x$ region, we have chosen to include them and determine our parton distributions together with the \mbox{twist-4} contributions; not to mention that the higher-twist contributions are of intrinsic interest themselves. Nevertheless one should keep in mind that there is an inherent uncertainty in any analysis due to data selection 
which ideally should be taken into consideration.
\bigskip

Finally it should be mentioned that we also performed a dynamical NNLO fit with a reduced cut $Q^2\geq 1$ GeV$^2$ (together with
$W^2\geq 3.5$ GeV$^2$ and twist-4 as in our nominal fit) and one with $Q^2\geq 3.5$ GeV$^2$, no twist-4 and no SLAC data, which are shown in Tab.~1.
The differences are rather small, even when compared with our nominal fit, as far as the final PDFs are concerned as well as
$\alpha_s(M_Z^2)$: in the first case $\alpha_s=0.1136\pm 0.0004$ and in the second one $\alpha_s=0.1143\pm 0.0004$
(cf.~Tab.~2 below).  This indicates that both choices, as well as our nominal one, are robust.
\bigskip

As stated in the Introduction, and in contrast to most other analyses, we do not assume a unique fixed value for the input scale $Q_0^2$, but rather pay special attention to the dependence of the results on this choice. Specifically we have considered systematic variations of this quantity ranging from 0.6 to 9 GeV$^2$. Variations of the input scale provide information on the relative size of the so-called procedural bias, i.e. the inability of the estimation procedure to find the optimal solution, for example shortcomings of the parametrization to reproduce the optimal shape of the distributions at the different input scales, as well as of the theoretical framework and the statistical estimation procedure \cite{ref13}. This is especially meaningful in the low $Q^2$ region, in which the low-$x$ gluon and sea distributions go through a complete rearrangement of their shape, changing \emph{naturally} from a valencelike structure (or even negative values) to a more conventional standard shape where they 
increase as $x$ decreases. Thus in addition to the systematic variations we will focus our attention on particular instances of dynamical ($Q_0^2 = 0.8 \;{\rm GeV}^2$) and standard ($Q_0^2 = 2 \;{\rm GeV}^2$) results, as in our previous analyses \cite{ref14,ref2}.
\bigskip

Requiring a valencelike behavior of the gluon input distribution, our fits favor a dynamical input scale of 0.8 GeV$^2$.  Smaller input scales imply increasing values of $\chi^2$ and decreasing values of $\alpha_s$ according to Figs.~1 and 2, respectively,
which is indicative for the increasing importance of nonperturbative contributions (higher twists).
In order not to deteriorate the quality of the fit and to avoid any bias towards low $\alpha_s$ values, we fix the dynamical input scale
to be $Q_0^2=0.8$ GeV$^2$ where the sea increases already slightly at small $x$.  Since the valencelike input gluon distribution
dominates the small-$x$ behavior of structure functions, we nevertheless refer to this solution as dynamical. Note that the distinction between dynamical and standard solutions is meaningful due to their different qualitative (physical) behavior, and is not only driven by the marginal $\chi^2$ dependence, i.e., since the value of $\chi^2$ for $Q_0^2=0.8$ GeV$^2$ in Fig.~1 is only marginally larger than for 2 GeV$^2$, the dynamical approach can still be pursued.
\bigskip

On the other hand we have fixed the standard input scale to be 2 GeV$^2$, as has become common by now, in order not to lose
valuable information for fixing PDFs in particular in the large-$x$ region, as discussed above, as well as to
avoid sizable backward evolutions.
This specific choice of the standard input scale is of minor importance, since above about 1.5 GeV$^2$ the final $\chi^2$
and $\alpha_s$ values in Figs.~1 and 2 are practically constant.  Furthermore the (incorrect) inclusion of jet (and charm) data
has little influence on our nominal NNLO results for $\alpha_s$ in Fig.~2 as shown by the NNLO* values at $Q_0^2=0.8$ and 
2 GeV$^2$.
\bigskip

For each (fixed) choice of the input scale we parametrize the input parton distributions $u_v=u-\bar{u}$, $d_v=d-\bar{d}$, $g$, $\Sigma=\bar{u}+\bar{d}$, $\Delta=\bar{d}-\bar{u}$, $s^+=s+\bar{s}$, and $s^-=s-\bar{s}$. The most general parametrization of the input parton distributions that we have considered in the current analyses can be written as
%equation 7
\begin{equation}
 \begin{array}{l}
xu_v(x,Q_0^2) = N_u x^{a_u}(1-x)^{b_u}(1 + A_u \sqrt{x} + B_u x + C_u x^2)\\[0.5em]
xd_v(x,Q_0^2) = N_d x^{a_d}(1-x)^{b_d}(1 + A_d \sqrt{x} + B_d x + C_d x^2)\\[0.5em]
xg(x,Q_0^2) = N_g\, x^{a_g}(1-x)^{b_g}(1 + B_g x^{\alpha_g}(1-x)^{\beta_g})\\[0.5em]
x\Sigma(x,Q_0^2) = N_\Sigma x^{a_\Sigma}(1-x)^{b_\Sigma}(1 + A_\Sigma \sqrt{x} + B_\Sigma x)\\[0.5em]
x\Delta(x,Q_0^2) = N_\Delta x^{a_\Delta}(1-x)^{b_\Delta}(1 + A_\Delta \sqrt{x} + B_\Delta x)\\[0.5em]
\tfrac{x}{2}s^+(x,Q_0^2) = N_{s^+} x^{a_{s^+}}(1-x)^{b_{s^+}}( 1 + A_{s^+} \sqrt{x} + B_{s^+} x)\\[0.5em]
xs^-(x,Q_0^2) = N_{s^-} x^{a_{s^-}}(1-x)^{b_{s^-}} (1 - \frac{x}{x_0})\\[0.5em]
 \end{array}
\end{equation}
although some of the parameters turn out to be superfluous and our nominal choice is somewhat simpler. The second term in parentheses in the gluon parametrization in principle allows for negative input gluons in the small-$x$ region. However, in all our fits the input gluon distribution remains positive, as was discussed in more detail in \cite{ref13}. As a matter of fact, if one chooses a sufficiently low input scale the details of the input distribution at low $x$ do not have much influence on the predictions at higher scales, which is one of the aspects exploited in the dynamical approach to parton distributions. We also tried to keep $B_g$ free with different (fixed) values for $\alpha_g$ and $\beta_g$ but found no significant improvement in $\chi^2$ accompanied by large correlations with other parameters; thus in our nominal parametrization we fix $B_g = 0$. Furthermore, because these
parameters did not provide a considerable improvement in the description of current data and/or in order to avoid flat directions in the parameter space (which would be associated with a singular error matrix), the following parameters were treated as follows: $B_\Delta = 0$, $b_{s^+}=b_\Sigma$, $A_{s^+}= A_\Sigma$, $B_{s^+}=B_\Sigma$, $a_{s^-} = 0.2$. Note as well that the usual quark number sum rules have been imposed, so that $N_u$, $N_d$ and $x_0$ in Eq.(7) are not free parameters, and similarly the momentum sum rule fixes $N_g$. This amounts to a total of 25 free parameters for the parton distributions at the input scale determined together with $\alpha_s(M_Z^2)$ and the 12 parameters of the higher-twist contributions (38 independent free parameters in total).
Our results are shown in Tables 2 and 3.
\bigskip

The typical dependence of $\chi^2$ on $\alpha_s(M_Z^2)$ for the various sets of data used is illustrated in Fig.~3 for our standard NNLO fit. Note that these curves represent the contributions to the total $\chi^2$ of the respective data subject to the constraints of \emph{all} data sets in the analysis, and thus the location of their minimum should not be strictly interpreted as the favored value by each data set (alone). Nevertheless it is interesting to see how the $\alpha_s(M_Z^2)$ dependence of the total $\chi^2$ arises. We find that many data sets are rather insensitive to $\alpha_s(M_Z^2)$ values; most other data favor values close to the minimum of the total $\chi^2$, while NMC (SLAC) tend to lower (higher) values. Because of the difficulties in the interpretation of such numbers, we refrain from providing preferred $\alpha_s(M_Z^2)$ values for individual data sets.  
\bigskip

%Table 2
\begin{table}[t!]
\centering
\begin{tabular}{|c|c c|c c|}
\multicolumn{1}{c}{}&\multicolumn{2}{c}{$Q_0^2 = 0.8 \;{\rm GeV}^2$}&\multicolumn{2}{c}{$Q_0^2 = 2 \;{\rm GeV}^2$}\\
\hline
                  &        NLO          &        NNLO         &        NLO          &        NNLO         \\
\hline
$\alpha_s(M_Z^2)$ & 0.1158 $\pm$ 0.0004 & 0.1136 $\pm$ 0.0004 & 0.1191 $\pm$ 0.0005 & 0.1162 $\pm$ 0.0006 \\
\hline
$a_u$             & 0.56   $\pm$ 0.03   & 0.71   $\pm$ 0.03   & 0.55   $\pm$ 0.02   & 0.71   $\pm$ 0.03   \\
$b_u$             & 3.51   $\pm$ 0.03   & 3.59   $\pm$ 0.05   & 3.61   $\pm$ 0.03   & 3.65   $\pm$ 0.07   \\
$A_u$             & 0.7    $\pm$ 0.5    &-0.2    $\pm$ 0.4    & 0.8    $\pm$ 0.4    &-0.8    $\pm$ 0.3    \\
$B_u$             & 6.4    $\pm$ 0.8    & 3.8    $\pm$ 0.5    & 4.7    $\pm$ 0.7    & 3.4    $\pm$ 0.3    \\
$C_u$             & 1.4    $\pm$ 0.8    & 0      $\pm$ 0.5    &-0.1    $\pm$ 0.6    &-1.3    $\pm$ 0.3    \\
\hline
$a_d$             & 0.87   $\pm$ 0.07   & 1.30   $\pm$ 0.06   & 0.92   $\pm$ 0.06   & 1.10   $\pm$ 0.05   \\
$b_d$             & 4.1    $\pm$ 0.3    & 4.9    $\pm$ 0.3    & 4.6    $\pm$ 0.3    & 5.0    $\pm$ 0.3    \\
$A_d$             &-1.9    $\pm$ 0.4    &-3.2    $\pm$ 0.2    &-2.8    $\pm$ 0.3    &-3.1    $\pm$ 0.2    \\
$B_d$             & 4.4    $\pm$ 0.6    & 4.1    $\pm$ 0.2    & 4.5    $\pm$ 0.4    & 4.2    $\pm$ 0.2    \\
$C_d$             &-2.4    $\pm$ 0.6    &-1.5    $\pm$ 0.3    &-2.0    $\pm$ 0.4    &-1.7    $\pm$ 0.4    \\
\hline
$a_g$             & 0.59   $\pm$ 0.06   & 0.91   $\pm$ 0.09   & 0.047  $\pm$ 0.014  & 0.123  $\pm$ 0.017  \\
$b_g$             & 8.3    $\pm$ 0.4    &12.0    $\pm$ 0.7    & 6.1    $\pm$ 0.2    & 7.5    $\pm$ 0.4    \\
\hline
$N_\Sigma$        & 0.076  $\pm$ 0.007  & 0.323  $\pm$ 0.028  & 0.164  $\pm$ 0.009  & 0.284  $\pm$ 0.012  \\
$a_\Sigma$        &-0.214  $\pm$ 0.009  &-0.070  $\pm$ 0.010  &-0.190  $\pm$ 0.007  &-0.150  $\pm$ 0.005  \\
$b_\Sigma$        & 7.91   $\pm$ 0.13   & 9.12   $\pm$ 0.15   & 8.42   $\pm$ 0.11   & 9.13   $\pm$ 0.10   \\
$A_\Sigma$        & 7.8    $\pm$ 0.9    &-2.0    $\pm$ 0.4    & 1.9    $\pm$ 0.3    &-1.0    $\pm$ 0.2    \\
$B_\Sigma$        & 21.8   $\pm$ 1.3    & 14.4   $\pm$ 0.8    & 10.0   $\pm$ 0.8    & 10.1   $\pm$ 0.3    \\
\hline
$N_\Delta$        & 57     $\pm$ 10     & 28     $\pm$ 6      & 37     $\pm$ 8      & 13     $\pm$ 3      \\
$a_\Delta$        & 2.29   $\pm$ 0.07   & 3.30   $\pm$ 0.07   & 2.20   $\pm$ 0.06   & 1.99   $\pm$ 0.05   \\
$b_\Delta$        & 18.6   $\pm$ 0.9    & 19.2   $\pm$ 0.7    & 19.2   $\pm$ 0.5    & 19.0   $\pm$ 0.4    \\
$A_\Delta$        & 1.0    $\pm$ 1.0    & 5.1    $\pm$ 2.0    & 2.1    $\pm$ 0.9    & 5.2    $\pm$ 1.2    \\
\hline
$N_{s^+}$         & 0.014  $\pm$ 0.003  & 0.081  $\pm$ 0.014  & 0.030  $\pm$ 0.004  & 0.058  $\pm$ 0.008  \\
$a_{s^+}$         &-0.12   $\pm$ 0.05   & 0.08   $\pm$ 0.05   &-0.28   $\pm$ 0.03   &-0.18   $\pm$ 0.03   \\
\hline
$N_{s^-}$         &-0.007  $\pm$ 0.008  & -0.006 $\pm$ 0.006  & -0.006 $\pm$ 0.005  &-0.005  $\pm$ 0.005  \\
$b_{s^-}$         & 18     $\pm$ 24     & 15     $\pm$ 10     &   15   $\pm$ 6      & 14     $\pm$ 11     \\
\hline
\end{tabular}
\caption{Values and experimental uncertainties (corresponding to the tolerance $\Delta\chi^2=1$) of the free parameters in Eq.(7) 
obtained in our nominal ``dynamical'' ($Q_0^2 = 0.8 \;{\rm GeV}^2$) and ``standard'' ($Q_0^2 = 2 \;{\rm GeV}^2$) NLO and NNLO analyses. The correlation coefficients and parameter values of the eigenvector sets are available from our web page \cite{ref55}.}
\label{parametersLT}
\end{table}                          

\begin{table}[t!]
\centering
\begin{tabular}{|c|c c|c c|}
\multicolumn{1}{c}{}&\multicolumn{2}{c}{$Q_0^2 = 0.8 \;{\rm GeV}^2$}&\multicolumn{2}{c}{$Q_0^2 = 2 \;{\rm GeV}^2$}\\
\hline
              &       NLO       &       NNLO      &        NLO      &       NNLO             \\
\hline
$h_2^p(0.1)$ &-0.045 $\pm$ 0.005 &-0.042 $\pm$ 0.006 &-0.033 $\pm$ 0.005 &-0.032 $\pm$ 0.006 \\
$h_2^p(0.3)$ &-0.049 $\pm$ 0.004 &-0.036 $\pm$ 0.004 &-0.054 $\pm$ 0.005 &-0.037 $\pm$ 0.005 \\
$h_2^p(0.5)$ & 0.025 $\pm$ 0.004 & 0.024 $\pm$ 0.004 & 0.007 $\pm$ 0.005 & 0.008 $\pm$ 0.005 \\
$h_2^p(0.7)$ & 0.041 $\pm$ 0.003 & 0.033 $\pm$ 0.003 & 0.031 $\pm$ 0.003 & 0.024 $\pm$ 0.003 \\
\hline
$h_2^n(0.1)$ &-0.050 $\pm$ 0.006 &-0.041 $\pm$ 0.007 &-0.041 $\pm$ 0.007 &-0.031 $\pm$ 0.007 \\
$h_2^n(0.3)$ &-0.012 $\pm$ 0.007 &-0.002 $\pm$ 0.007 &-0.013 $\pm$ 0.007 &-0.003 $\pm$ 0.007 \\
$h_2^n(0.5)$ & 0.020 $\pm$ 0.006 & 0.013 $\pm$ 0.006 & 0.007 $\pm$ 0.006 & 0.004 $\pm$ 0.006 \\
$h_2^n(0.7)$ & 0.016 $\pm$ 0.004 & 0.015 $\pm$ 0.004 & 0.014 $\pm$ 0.004 & 0.012 $\pm$ 0.004 \\
\hline
$h_L^p(0.2)$ & 0.016 $\pm$ 0.012 &-0.000 $\pm$ 0.012 & 0.005 $\pm$ 0.012 &-0.014 $\pm$ 0.010 \\
$h_L^p(0.6)$ & 0.035 $\pm$ 0.006 & 0.027 $\pm$ 0.006 & 0.031 $\pm$ 0.006 & 0.024 $\pm$ 0.006 \\
\hline
$h_L^n(0.2)$ &-0.002 $\pm$ 0.020 &-0.009 $\pm$ 0.021 &-0.004 $\pm$ 0.021 &-0.014 $\pm$ 0.021 \\
$h_L^n(0.6)$ &-0.005 $\pm$ 0.011 &-0.003 $\pm$ 0.011 &-0.001 $\pm$ 0.011 &-0.002 $\pm$ 0.011 \\
\hline
\end{tabular}
\caption{As in Table \ref{parametersLT} but for the twist-4 coefficients in Eq.(3) in units of GeV$^2$.}
\label{parametersHT}
\end{table}

Our standard and dynamical NNLO results for the non-singlet and singlet PDFs at $Q^2=10$ GeV$^2$ are compared in Figs.~4 and 5
with our previous dynamical JR09 ones \cite{ref2} as well as with the ones of MSTW08 \cite{ref56} and ABM11 \cite{ref5}.
(The most recent global ABM12 fit \cite{ref57} takes into account LHC data as well but the results turn out to be rather similar
to the ones of ABM11.)
Our strange sea asymmetry ($s-\bar{s}\neq 0$) in Fig.~4 is now compatible with other determinations which is in contrast to
our previous JR09 result \cite{ref2} where $s=\bar{s}$ has been assumed.
Similarly, our new strange sea distribution $s+\bar{s}$ in Fig.~5 lies in the right ballpark, while the discrepancy of the much smaller
JR09 result is traced back to our previously assumed input ansatz $s(x,Q_0^2) =\bar{s}(x,Q_0^2)=0$  which implied a purely
dynamically generated strange sea at $Q^2>Q_0^2$.
On the other hand, both new JR14 gluon distributions in Fig.~5 are compatible with the previous JR09 one which, in the medium-$x$
region, lie between the ones of ABM11 and MSTW08.  
At small $x$, however, our gluon densities are in agreement with ABM11, taking into account that the dynamical gluon densities 
are steeper than the standard ones due to the longer $Q^2$-evolution starting at $Q_0^2=0.8$ GeV$^2$, which also implies
stronger constrained (smaller errors) distributions.
These features become even more transparent by plotting the respective ratios of the PDFs as done in Fig.~6.  Furthermore,
the differences between the NNLO and NLO distributions are depicted in Fig.~7 by the NNLO/NLO ratios of our dynamical solution.
In addition, our new (dynamical) NLO PDFs are, apart from the strange sea density, similar to our previous (dynamical) NLO GJR
ones {\cite{ref14}: 
at $Q^2=10$ GeV$^2$, for example, the GJR gluon is somewhat larger (by about 15\% at $x\simeq 10^{-5}$, comparable to
JR14 at $x=10^{-3} - 0.1$, and sizably larger at large $x$ since the JR14 gluon decreases much faster as $x\to 1$); the $u+\bar{u}$
density is rather similar everywhere, which holds also for $d+\bar{d}$ except in the large-$x$ region where the GJR $d+\bar{d}$ 
becomes about 30\% smaller at $x\simeq 0.7$ since it decreases faster than the present one.
\bigskip

It should be noted that at a scale $Q^2 = 10^4$ GeV$^2$, relevant for gauge- and Higgs-boson as well as top-quark production at hadron colliders, the differences between the various sets of parton distributions in Figs. 4 -- 6 become less pronounced with somewhat reduced error bands, in particular in the small-x region.

The twist-4 contributions to the structure functions in Eq.~(3) are shown in Fig.~8 which are similar in shape and size as the ones
obtained by ABM11 \cite{ref5}.  Typically, the contributions to $F_2$ turn negative below about $x\simeq$ 0.5 where they remain sizable
and non-negligible even at moderately small values of $x$. The higher-twist contributions to the proton longitudinal structure function 
are smaller at moderate $x$ values but turn significantly positive at larger $x\gtrsim 0.5$, where their values are driven by JLab and SLAC data, remarkably by the $F_2/F_L$ separated SLAC-E140x measurements \cite{ref42}. Our results for $F_L^n$ are compatible with zero within present experimental uncertainties. The inclusion of these data
at large $x$ is a novelty of our global analysis and has also resulted in a sizable reduction of the experimental uncertainties on the gluon distribution at large $x$, as can be seen in Fig.~\ref{ratios} (note in addition that the JR09 uncertainties \cite{ref2} correspond to $\Delta \chi^2 > 1$). For completeness we compare in Fig.~9 our dynamical and standard predictions for $F_L(x,Q^2)$ with recent HERA(H1) data \cite{ref41}.  It should be noticed that these theoretical and experimental results cover a very large domain of $x$
which ranges from $x=2.8 \times 10^{-5}$ (at $Q^2=1.5$ GeV$^2$) to $x =1.84\times 10^{-2}$ (at $Q^2=636$ GeV$^2$). Our predictions for the different proton and neutron structure functions can be obtained in \cite{ref55}.
\bigskip

The HERA charm production data \cite{ref40} are compared with our results in Fig.~10.  As already discussed, we use these
data only for the NLO fits due to our ignorance of the massive NNLO partonic sub-cross sections.  Our dynamical NLO results
are in good agreement with experiment ($\chi^2=1.41$ per data point, according to Tab.~1).  
In the standard case $\chi^2$ is generally worse (cf.~Tab.~1) due to the correlation between these data and the value of
$\alpha_s$:  the data prefer smaller $\alpha_s$ values and in our case the standard fits generally tend towards larger values.
The dynamical NNLO* fit shown in Fig.~10 corresponds to $\chi^2=1.68$ (see Tab.~1), but such a comparison might be 
deceptive since the order of the PDFs appears to be less relevant than the correct (unknown) NNLO matrix elements \cite{ref53}.
(If we just take our dynamical NNLO PDFs and combine them with the massive NLO Wilson coefficients (no fit), we obtain
$\chi^2=1.93$ for the charm data in Fig.~10.)  
Finally, we compare \mbox{the charm} data with approximate NNLO$_{\rm approx}$ expectations as derived in \cite{ref37}; 
such corrections to massive 3-loop Wilson coefficients are obtained by interpolating between existing soft-gluon threshold
resummation results and asymptotic ($Q^2\gg m_h^2$) coefficients \cite{ref58,ref59}.  
The interpolation uncertainty is quantified by two options \cite{ref37}, $c_2^{(2),A}$ and $c_2^{(2),B}$, for the constant
terms in the NNLO correction as a linear combination of these two options,
%%%%%%%%%%%%%%%%%%%%%%%%%%%%%%%%%%%%%%%%%%%%%%%%%%%%%%
%Equation(8)
\begin{equation}
c_2^{(2)} = (1-d_N)c_2^{(2),A} + d_N\, c_2^{(2),B}
\end{equation}
%%%%%%%%%%%%%%%%%%%%%%
with the interpolation parameter $d_N$ to be fitted to the data.  Our optimal NNLO$_{\rm approx}$ predictions in Fig.~10
correspond to $d_N=-0.62\pm 0.09$ (corresponding to $\chi^2=1.30$) which should be compared with \cite{ref60}
$d_N=-0.4$ based on the ABM11 PDFs, and \cite{ref57} $d_N=-0.1\pm 0.15$ based on the ABM12 PDFs.
(Choosing $d_N=0$ would give $\chi^2= 2.10$ instead of the optimal $1.30$.)  This indicates a strong dependence
of $d_N$ on the specific PDF sets used.  Anyway, the dynamical and standard NLO predictions and the various NNLO
expectations in Fig.~10 are in good agreement with experiment and are practically indistinguishable. This demonstrates
that the in principle unambiguous fixed-order perturbative predictions for heavy-quark production within the 3-flavor
FFNS are sufficient for describing and explaining present experiments \cite{ref5,ref10,ref14,ref57,ref61,ref62}.
\bigskip

For the time being one cannot consistently use hadronic and DIS jet production measurements for NNLO analyses 
because the required NNLO matrix elements are not fully known yet (the calculation of these full NNLO QCD corrections
is still in progress \cite{ref52}).
This is unfortunate since jet data are sensitive to all PDFs of the nucleon and provide the most direct constraint on
the gluon distribution at large $x$.  Apart from an entirely consistent NLO {\mbox analysis} we nevertheless perform a NNLO
fit using NLO partonic sub-cross sections (referred to as NNLO*, cf.~Tab.~1) in order to check any potential impact
of the jet data on our NNLO PDFs and $\alpha_s$.  
However, one should keep in mind that the correct order of matrix elements would be far more important than the
correct order of PDFs \cite{ref53}.  
In Fig.~11 we compare some typical representative hadronic jet-data sets with the theoretically fully consistent NLO
results (the dynamical ones shown are of similar quality as the standard ones, cf.~Tab.~1) as well as with the NNLO* 
fit and with NNLO (where jet cross sections are calculated, not fitted, as in NLO but just using our fixed NNLO PDFs).
The jet data are already reasonably well described at NLO, and the differences between NLO and the theoretically
inconsistent NNLO* and NNLO results are small.  The same holds for the HERA jet data.  This again illustrates that
the incorrect order of the PDFs is not very important \cite{ref53}.  Furthermore, the jet data have little influence on
the consistently determined NNLO value of $\alpha_s$ (cf.~Tab~1 and 2 where no jet data have been used) which
is illustrated by the NNLO* $\alpha_s(M_Z^2)$ values depicted in Fig.~2.  
In the standard fit the NNLO $\alpha_s(M_Z^2)$ is even practically unaltered.  This confirms results already obtained
and emphasized in \cite{ref5,ref57,ref63}, in contrast to claims made by other PDF groups (for a comparative review,
see \cite{ref64}, for example).  A recent summary and critical discussion of current $\alpha_s$ results can be found
in\cite{ref57}.

%%%%%%%%%%%%%%%%%%%%%%%%%%%%%%%%%%%%%%%%%%%%%%%%%%%%%%%%%%%%%%%
\section{Predictions for Hadron Colliders}
Finally we turn to our {\em{pre}}dictions for hadron colliders.  As emphasized in the Introduction, we did not include
TeVatron gauge-boson production data and LHC data in our fitting procedure in order to allow for genuine predictions
of these measurements.  Thus one can explicitly test the reliability and usefulness of the QCD-improved parton model
and of QCD in general.
The general theoretical framework for hadronic weak gauge-boson ($W^{\pm}, Z^0$) and SM Higgs-boson ($H^0$)
production up to NNLO has been summarized in \cite{ref3}, for example, and will not be repeated here.
\bigskip

Our predictions for gauge-boson production at the TeVatron ($\sqrt{s}=1.8$ TeV) are given in Tab.~4 where they 
are compared with our previous NNLO dynamical JR09 \cite{ref3} expectations as well as with the ones of ABM11
\cite{ref5} and of MSTW08 \cite{ref1,ref56}.  
We confirm the slightly enhanced production rates of ABM11 which are, within $1\sigma$, also in agreement with
experiment.  
The predictions for LHC at $\sqrt{s}=7,8$, and 14 TeV are presented in Tables 5, 6, and 7, respectively. 
It should be noted that the $W^\pm$ and $Z^0$ cross sections are very highly correlated, so that their ratio depends
very little on specific PDFs, while the ratio of $W^+/W^-$ cross sections is a sensitive probe of the $u/d$ ratio (see,
e.g., \cite{ref11} for more details).
To illustrate these correlations we show in Fig.~12 $W^{\pm}\equiv W^+ +W^-$ versus $Z^0$ and $W^+$
versus $W^-$ total cross sections by drawing ellipses to account for the correlations between the two cross sections,
both for the experimental measurements and for the theoretical predictions.  The reduction of uncertainty in the
$W^+/W^-$ cross-section ratio (cf.~Tables 5 -- 7) is seen as a shrinking of the corresponding ellipse, due to an 
improved knowlegdge of the light-quark flavor separation. 
It should be furthermore emphasized that our new dynamical and standard predictions in Tab.~6 are in excellent
agreement with most recent CMS measurements at $\sqrt{s}=8$ TeV \cite{ref69}.
\bigskip
 
%%%%%%%%%%%%%%%%% Table 4 %%%%%%%%%%%%%%%%%%%%%
\begin{table}[t!]
\centering
\begin{tabular}{c|c|c |c|c||c|c|}
\multicolumn{4}{c}{} & \multicolumn{1}{c||}{} & \multicolumn{2}{c|}{data} \\ \hline
                           & dynJR14 & stdJR14  & ABM11 & MSTW08 & CDF & D0\\
\hline
$W^{\pm}$  & 24.07 $\pm$ 0.14  & 24.37 $\pm$ 0.14  & 24.2 $\pm$ 0.3  & 23.14 $\pm$ 0.39 & 23.16 $\pm $1.12 & 21.95 $\pm$ 1.40 \\
& (23.07 $\pm$ 0.24)    &    &  &  &  &  \\
$Z^0$   & 7.30 $\pm$ 0.04   & 7.38 $\pm$ 0.04  & 7.28 $\pm$ 0.11  & 6.77 $\pm$ 0.13 &  6.87 $\pm$ 0.36 &  6.48 $\pm$ 0.48\\
& (6.98 $\pm$ 0.07)    &    &  &  & & \\
\hline
\end{tabular}
\caption{NNLO predictions, together with the $1\sigma$ PDF uncertainties, for the production cross sections $\sigma(p\bar{p}\to V+X)\, [nb]$ at $\sqrt{s}=1.8$ TeV with $V=W^{\pm},Z^0$ where $W^{\pm}\equiv W^+ +W^-$.
Note that for $p\bar p$ collisions the $W^+$ and $W^-$ cross sections are equal.  For 
comparison, our previous dynamical results \cite{ref2,ref3} are shown in parentheses, and the standard
ABM11 \cite{ref5} and MSTW08 \cite{ref1,ref56} expectations are presented as well.  The TeVatron
data are taken from \cite{ref65,ref66}.}
\end{table}

%%%%%%%%%%%%%%%%%% Table 5 %%%%%%%%%%%%%%%%%%%%
\begin{table}[t!]
\centering
\begin{tabular}{l|c|c|c|c|}
     & dynJR14 (JR09) & stdJR14  & ABM11 &  MSTW08 \\
\hline
$W^+$       &  56.9 $\pm$ 0.4 (54.6 $\pm$  1.1) & 57.9 $\pm$ 0.4 & 59.5 $\pm$  0.9   & 56.8 $\pm$  1.0\\
$W^-$        &  39.7 $\pm$ 0.3 (37.2 $\pm$  0.8) & 40.4 $\pm$ 0.3 & 40.0 $\pm$  0.7   &  39.6 $\pm$  0.7\\
$W^{\pm}$ &  96.6 $\pm$ 0.6 (91.7 $\pm$ 1.8) &  98.4 $\pm$ 0.7 & 99.5 $\pm$  1.4   &  96.4 $\pm$  1.6 \\
$Z^0$         &  28.8 $\pm$ 0.2 (27.2 $\pm$ 0.5) &  29.4 $\pm$ 0.2 & 29.2 $\pm$  0.4   &  27.9 $\pm$  0.5\\
\hline
$W^+/W^-$       &  1.443 $\pm$  0.005 &  1.433 $\pm$  0.005 & -- & --\\
$W^{\pm}/Z^0$  &  3.351 $\pm$  0.040 &  3.350 $\pm$ 0.004 &  -- & --\\
$W^+/Z^0$        &  1.974 $\pm$  0.004 &  1.973 $\pm$ 0.004 &  -- & --\\
$W^-/Z^0$         &  1.377 $\pm$  0.003 &  1.377 $\pm$ 0.003 &  -- &  --\\
\hline
\end{tabular}
\caption{NNLO predictions for the production cross sections $\sigma(pp\to V+X)\, [nb]$ at
$\sqrt{s}=7$ TeV with $V=W^{\pm},\, Z^0$ where $W^{\pm}\equiv W^++W^-$.  
To allow for a comparison with previous results we also list the corresponding cross sections
of the dynamical JR09 \cite{ref2,ref3} in parentheses, as well as the ones of ABM11 \cite{ref5}
and MSTW08 \cite{ref1,ref56}. The errors refer to the dominant $1\sigma$ PDF uncertainties.}
\end{table}

%%%%%%%%%%%%%%%%%% Table 6 %%%%%%%%%%%%%%%%%%%%
\begin{table}[t!]
\centering
\begin{tabular}{l|c|c|c|c|}
     & dynJR14 (JR09) & stdJR14  & ABM11 &  MSTW08 \\
\hline
$W^+$  &      65.4 $\pm$  0.4 (62.6 $\pm$ 1.4)  & 66.5 $\pm$  0.5 &  68.3 $\pm$  1.0 & 65.4 $\pm$  1.1  \\
$W^-$   &      46.2 $\pm$  0.3 (43.3 $\pm$ 1.0)  & 47.1 $\pm$  0.4 &  46.7 $\pm$  0.8 &  46.3 $\pm$  0.7\\
$W^{\pm}$ & 111.6 $\pm$  0.7 (105.9 $\pm$ 2.3) & 113.6 $\pm$ 0.8  &115.0 $\pm$ 1.7 & 111.7$\pm$  1.8 \\
$Z^0$         & 33.5 $\pm$  0.2 (31.6 $\pm$ 0.6) & 34.1 $\pm$ 0.2 &  34.0 $\pm$ 0.5 & 33.5 $\pm$ +0.5 \\
\hline
$W^+/W^-$       &   1.413 $\pm$ 0.005 & 1.413 $\pm$ 0.005 &  -- & --\\
$W^{\pm}/Z^0$  &  3.332 $\pm$  0.004 & 3.330 $\pm$ 0.004 & -- & --\\
$W^+/Z^0$        &  1.951 $\pm$  0.004 & 1.950 $\pm$ 0.004 &  -- & --\\
$W^-/Z^0$         &  1.381 $\pm$  0.003 &  1.380 $\pm$ 0.003 &  -- & -- \\
\hline
\end{tabular}
\caption{As in Table 5 but for $\sqrt{s}=8$ TeV.}
\end{table}

%%%%%%%%%%%%%%%%%% Table 7 %%%%%%%%%%%%%%%%%%%%
\begin{table}[t!]
\centering
\begin{tabular}{l|c|c|c|c|}
     & dynJR14 (JR09) & stdJR14  & ABM11 &  MSTW08 \\
\hline
$W^+$        &    114.6 $\pm$ 0.8 (109.3 $\pm$ 3.1)  & 116.4 $\pm$ 0.9  & 119.0 $\pm$ 1.8  & 114.0 $\pm$ 2.0 \\
$W^-$         &    85.4 $\pm$ 0.7 (80.0 $\pm$ 2.3) &  86.9 $\pm$ 0.7 &  86.6 $\pm$ 1.4 & 85.6 $\pm$ 1.5\\
$W^{\pm}$  &  200.0 $\pm$  1.4 (189.3 $\pm$ 5.4) & 203.3 $\pm$  1.6 & 205.7 $\pm$ 3.1 & 199.6 $\pm$  3.4\\
$Z^0$         &  61.4 $\pm$  0.4 (57.9 $\pm$ 1.6)&  62.5 $\pm$ 0.5 & 62.3 $\pm$  1.0 & 59.0 $\pm$ 1.0\\
\hline
$W^+/W^-$       &  1.342 $\pm$  0.004 &  1.340 $\pm$  0.004 & --  & --\\
$W^{\pm}/Z^0$  &  3.259 $\pm$ 0.004 &  3.255 $\pm$  0.004 &  -- & --\\
$W^+/Z^0$        &  1.867 $\pm$  0.003 & 1.864 $\pm$  0.004 &  -- & --\\
$W^-/Z^0$         &  1.392 $\pm$  0.002 &  1.391 $\pm$ 0.002 &  -- & -- \\
\hline
\end{tabular}
\caption{As in Table 5 but for $\sqrt{s}= 14$ TeV.}
\end{table}

%%%%%%%%%%%%%%%%%%%%%%%%%%%%%%%%%%%%%%%%%%%%%

The NNLO predictions for SM Higgs-boson production at the LHC for present and future energies are given in Tab.~8 
corresponding to a Higgs mass $M_H = 125.5$ GeV \cite{ref70,ref71}.
The theoretical uncertainty due to the scale variation dominates by far over the $1\sigma$ PDF uncertainty.  
We observe rather small changes with respect to our previous dynamical JR09 \cite{ref3} results; the same holds true
for the standard ones in Tab.~8 which are even closer to the ABM11 \cite{ref5} predictions.  
A similar stability has been observed by ABM12 \cite{ref57} where LHC data have been included in the fits.
Interestingly, these predicted production rates are lower by some 10\% than the ones recommended for ongoing
ATLAS and CMS analyses by the HiggsXSWG \cite{ref72} which read
$15.13_{-1.18-1.07}^{+1.07+1.15}$pb and $19.27_{-1.50-1.33}^{+1.39+1.45}$pb at $\sqrt{s}=7$ and 8 TeV,
respectively; the second PDF uncertainties appear to be unreasonably large as compared to the $1\sigma$ PDF
uncertainties in Tab.~8.  For completeness we also display our predictions for Higgs-boson production at the TeVatron
in Tab.~9.  Our new results are, within uncertainties, comparable with the ones of other groups, although the less
recent MSTW08 expectations appear to be slightly enhanced.
\bigskip

We close this Section with a few comments concerning the top-quark pair production at the LHC, since the calculations
of the NNLO QCD corrections to hadronic $t\bar{t}$ production have been recently completed \cite{ref73,ref74,ref75}.
For a pole mass $m_t=173$ GeV we obtain, according to our dynamical
and standard PDFs, at $\sqrt{s}=7$ TeV
%%Equation 9 + 10 %%%%
\begin{eqnarray}
\sigma_{t\bar{t}}^{\rm dyn} & = & 143.2_{-5.8}^{+5.4} \pm 2.4\,  \rm{pb}\, \\
\sigma_{t\bar{t}}^{\rm std} & = & 154.1_{-6.5}^{+6.1} \pm 3.0\, \rm{pb}
\end{eqnarray}
%%%%%%%%%%%%%%
where the central values refer to a scale choice $\mu_F = \mu_R = m_t$.  The scale uncertainties are due to
varying $\mu_F=\mu_R$ between $m_t/2(+)$ and $2 m_t(-)$.  The second errors refer to the $\pm 1\sigma$ PDF
uncertainties.  For comparison, our previous dynamical JR09 \cite{ref3} PDFs predict $164.1 \pm 12.4$ pb
which is somewhat larger than in Eq.~(9) due to the higher and less-constrained JR09 gluon distribution in the small-$x$
region as shown in Fig.~6; the same holds for the standard prediction in Eq.~(10).
  These results are comparable to the ones obtained by
other PDF groups \cite{ref11,ref57} which should be compared to the measured cross section
$\sigma_{t\bar{t}}=173.3 \pm 10.1$\, pb \cite{ref76,ref77,ref78}.
\bigskip

For TeVatron measurements at $\sqrt{s}=1.96$ TeV we predict
%%Equation 11 + 12 %%%%
\begin{eqnarray}
\sigma_{t\bar{t}}^{\rm dyn} & = & 7.07_{-0.19}^{+0.22} \pm 0.06\,  \rm{pb}\, \\
\sigma_{t\bar{t}}^{\rm std} & = & 7.37_{-0.21}^{+0.25} \pm 0.07\, \rm{pb}
\end{eqnarray}
%%%%%%%%%%%%%%
whereas our previous dynamical JR09 \cite{ref3} PDFs give $7.1\pm 0.4$ pb.  The combined (CDF, D0) 
TeVatron top-quark pair cross section is measured to be $7.60\pm 0.41$ pb \cite{ref78,ref79}.

%%%%%%%%%%%%%%%%%% Table 8 %%%%%%%%%%%%%%%%%%%%
\begin{table}[t!]
\centering
\begin{tabular}{l|c|c|c|c|}
 $\sqrt{s}$/TeV    & dynJR14 (JR09) & stdJR14  & ABM11 &  MSTW08 \\
\hline
\,\,\,7    & $12.81_{-1.10}^{+1.28} \pm 0.14$ $(13.03_{-1.17}^{+1.24} \pm 0.41)$    
          &  $13.44_{-1.19}^{+1.41} \pm 0.16$  & $13.23_{-1.31}^{+1.35} \pm 0.30$ & $14.39_{-1.47}^{+1.54} \pm 0.20$\\
\,\,\,8    &  $16.39_{-1.35}^{+1.61} \pm 0.18$ $(16.55_{-1.44}^{+1.54} \pm 0.53)$  
         &  $17.18_{-1.47}^{+1.75} \pm 0.20$ & $16.99_{-1.63}^{+1.69} \pm 0.37$ & $18.36_{-1.82}^{+1.92} \pm 0.25$\\
14  &  $42.77_{-3.38}^{+3.81} \pm 0.41$ $(42.16_{-3.26}^{+3.60} \pm 1.59)$ 
         &  $44.64_{-3.32}^{+4.16} \pm 0.46$ &  $44.68_{-3.78}^{+4.02} \pm 0.85$  & $47.47_{-4.18}^{+4.52} \pm 0.62$ \\
\end{tabular}
\caption{NNLO predictions for the SM Higgs-boson production cross sections $\sigma(pp\to H^0+X)\, [pb]$ 
as a function of LHC energies via the dominant gluon-gluon fusion subprocess. For a comparison we also show our 
previous dynamical JR09 results \cite{ref2,ref3} and the ones of ABM11 \cite{ref5} and MSTW08 \cite{ref1,ref56}. 
The central values correspond to a scale choice $\mu_F=\mu_R=M_H$ with $M_H=125.5$ GeV.  The errors refer to
the scale uncertainties due to varying $\mu_F=\mu_R$ between $\frac{1}{2}
M_H(+)$ and $2M_H(-)$ and, respectively, the $\pm 1\sigma$ PDF uncertainties.}
\end{table}

%%%%%%%%%%%%%%%%%% Table 9 %%%%%%%%%%%%%%%%%%%%
\begin{table}[t!]
\centering
\begin{tabular}{l|c|c|c|c|}
$\sqrt{s}$/TeV & dynJR14 (JR09) & stdJR14  & ABM11 &  MSTW08 \\
\hline
1.8     &  $0.54_{-0.07}^{+0.07} \pm 0.012$ $(0.64_{-0.08}^{+0.09} \pm 0.05)$ 
          &  $0.59_{-0.08}^{+0.08} \pm 0.014$ & $0.50_{-0.07}^{+0.07} \pm 0.03$ & $0.67_{-0.10}^{+0.09} \pm 0.02$ \\
1.96    &  $0.69_{-0.08}^{+0.09} \pm 0.014$ $(0.81_{-0.10}^{+0.10} \pm 0.06)$
           &  $0.75_{-0.09}^{+0.10} \pm 0.017$ & $0.65_{-0.09}^{+0.09} \pm 0.04$ & $0.84_{-0.12}^{+0.12} \pm 0.03$ \\
\end{tabular}
\caption{As in Table 8 but for $\sigma (p\bar{p}\to H^0+X)\, [pb]$ as a function of TeVatron
energies.}
\end{table}

%%%%%%%%%%%%%%%%%%%%%%%%%%%%%%%%%%%%%%%%%%%%%%%%%%
\section{Summary and Conclusions}
Utilizing recent DIS measurements and data on Drell-Yan dilepton and (partly) high-$p_T$ inclusive jet production,
we have redone previous global fits \cite{ref2,ref14} for obtaining parton distributions of the nucleon of high statistical
accuracy, together with the highly correlated strong coupling $\alpha_s$, up to NNLO of perturbative QCD.  This has
been done within the so-called dynamical $(Q_0^2< 1$ GeV$^2$) and standard ($Q_0^2\gtrsim 1$ GeV$^2$) approaches,
with $Q_0$ being the (chosen) input scale where the RG $Q^2$-evolutions are started.
To ensure the reliability of our results we have now included nonperturbative higher-twist (HT) terms, and nuclear
corrections for the deuteron structure functions together with off-shell and nuclear shadowing corrections, as well
as target-mass corrections for $F_2$ and $F_L$.  To learn about the stability of the results, we performed fits to
subsets of data by applying various kinematic ($Q^2,W^2$) cuts on the available data, since in particular the HT
contributions turned out to be sensitive to such choices. 
A safe and stable choice turned out to be $Q^2\geq 2$ GeV$^2$ and $W^2\geq 3.5$ GeV$^2$, which are our
final nominal cuts imposed on DIS data.
In addition we studied the dependence of the results on the specific choice of $Q_0^2$ which in most cases has 
not been systematically addressed so far.
\bigskip

In addition to the above improvements in the theoretical computations as compared to our previous ones
\cite{ref2,ref3,ref14} is a complete treatment of the systematic uncertainties of the data including experimental
correlations.
\bigskip

Yet another improvement is the use of the running-mass definition for DIS charm and bottom production which results
in an improved stability of the perturbative series.  Since heavy quark coefficient functions are exactly known only 
up to NLO, we have refrained from using DIS heavy-quark production data for our nominal NNLO fits.  Nevertheless
the NLO results are in good agreement with experiment.  The same holds true when fitting charm data at NNLO
using (theoretically inconsistent) NLO matrix elements (referred to as NNLO*) or approximate NNLO ones.  
It should, however, be kept in mind that the correct order of (massive) matrix elements appears to be far more
important than the chosen order of PDFs \cite{ref53}.  
These results demonstrate that the, in principle, unambiguous quantum-field-theoretic fixed-order perturbative
predictions for heavy quark production within the fixed (three) flavor number scheme (FFNS), relying on three
light flavors in the initial state, are sufficient for describing and explaining present experiments.
\bigskip

We face a similar situation for DIS and hadronic inclusive jet production measurements where the required 
sub-cross sections are not fully known yet beyond NLO, although our NLO fits agree well with present experiments
(cf.~Fig.~11). In order to illustrate the effects of the incorrect inclusion of these data beyond NLO, we carried
out fits based on NNLO PDFs but using NLO matrix elements, again referred to as NNLO*.  The differences
between the NLO and the (theoretically inconsistent) NNLO* results turn out be small in the dynamical and
standard approaches (cf.~Fig.~11 and Tab.~1).  Again, the correct order of matrix elements would be far more
important \cite{ref53}.  Furthermore, the jet data have little influence on the consistently determined value of
$\alpha_s$; they leave $\alpha_s$ essentially unaltered in NNLO standard fits with $Q_0^2\gtrsim 1.5$ GeV$^2$
(cf.~Fig.~2).
\bigskip

The strong couplings obtained from our  nominal dynamical NNLO (NLO) analyses are $\alpha_s(M_Z^2) =
0.1136\pm 0.0004$ $(0.1158\pm 0.0004)$, whereas the somewhat less constrained standard fits give
$\alpha_s(M_Z^2)=0.1162\pm 0.0006$ $(0.1191\pm 0.0005)$. Note that the uncertainties quoted here are due only to the propagation of the experimental errors of the data included in the analysis (see also \cite{ref80}), while the differences between our dynamical and standard results have a systematic origin which is referred to as procedural bias. As a matter of fact it has been suggested \cite{ref13} that these differences can be used to estimate these systematic uncertainties; following the procedure proposed in \cite{ref13} an uncertainty of $\Delta_{\rm proc.}\alpha_s\!=\!0.0013$ should be attributed to procedural uncertainties only. Of course, this does not exhaust the list of systematic uncertainties, choices like data selection and genuinely theoretical uncertainties like scheme and scale choices in the analysis should also be considered, and would further increase the total error. A recent summary and comparative discussion of current $\alpha_s$ results can be found in \cite{ref57}.
\bigskip

It should be emphasized that, on purpose, we have not included TeVatron gauge-boson production data and
LHC data in our fitting procedure, in order to allow for genuine {\em predictions} of these measurements.
This allows us to explicitly test the reliability and usefulness of the QCD-improved parton model and QCD in
general.  Our previous dynamical (and standard) NNLO JR09 \cite{ref3} predictions for the TeVatron are 
similar to the present new ones, confirm the somewhat enhanced production rates of ABM11 \cite{ref5}, 
and are in agreement with experiment.  The same holds for our predictions for the LHC at present and future
energies ($\sqrt{s}=7,\, 8$, and 14 TeV) which are about 5\% larger than the previous JR09 ones \cite{ref3},
and compare now well with the ones of ABM11 \cite{ref5} and MSTW08 \cite{ref1,ref56}, for example.
The highly correlated $W^{\pm}$ and $Z^0$ cross sections are in good agreement with LHC measurements
at $\sqrt{s}=7$ TeV, and in particular our new dynamical and standard predictions are in excellent agreement
with most recent CMS measurements at $\sqrt{s}=8$ TeV \cite{ref69}.
\bigskip

Our new dynamical and standard NNLO predictions for Higgs-boson production at the LHC differ rather little from
our previous JR09 \cite{ref3} results and are close to the ones of ABM \cite{ref5,ref57} (cf.~Tab.~8).
Interestingly, these predicted rates are lower by some 10\% than the ones recommended for ongoing ATLAS
and CMS analyses \cite{ref72}.
\bigskip

Our dynamical and standard PDFs and predictions, the correlation coefficients and parameter values of the eigenvector sets can be obtained on request or directly from our web page \cite{ref55}.

\section*{\bf\Large Acknowledgements}
We thank S.~Alekhin and J. Bl\"umlein for several discussions and helpful comments, as well as M. Gl\"uck for a clarifying suggestion. One of us (P.J.-D.) is also indebted to W.~Melnitchouk, A.~Accardi and C.E.~Keppel for discussions. The work of P.J.-D. was supported by DOE contract
No. DE-AC05-06OR23177, under which Jefferson Science Associates, LLC operates Jefferson Lab.

%%%%%%%%%%%%%%%%%%%%%%%%%Figures%%%%%%%%%%%%%%%%%%%%%%%%%%%%%%

%%%%%%%%%%%%%%%Figure 1  ("chi2")
\begin{figure}[f]
\begin{center}
\includegraphics[width=\textwidth]{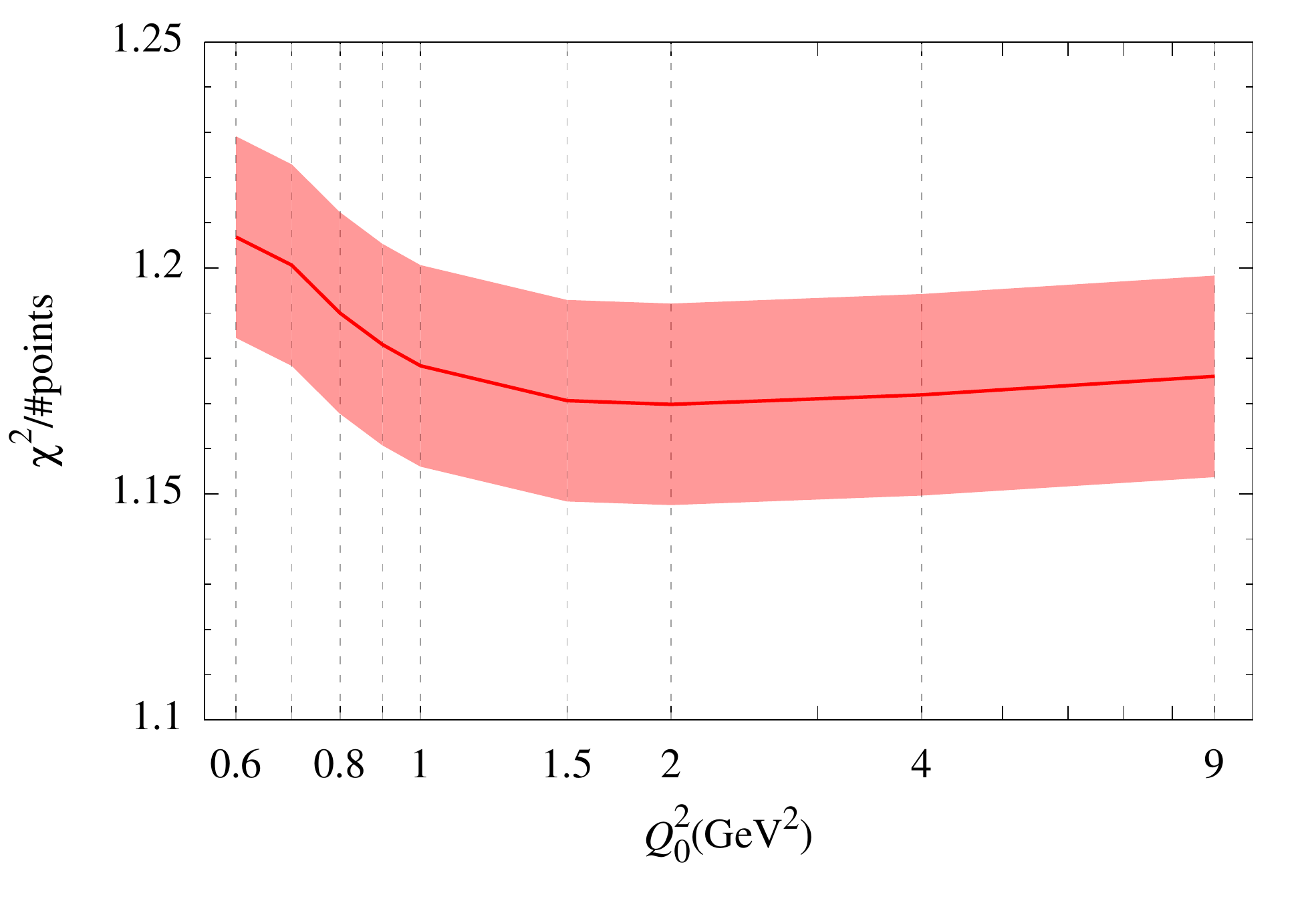}
\caption{The dependence of $\chi^2$ on the variations of the input scale $Q_0$ as obtained in our NNLO analyses.
The band indicates the $\pm 1\sigma$ uncertainties ($\Delta\chi^2\simeq \sqrt{2N}$ for \mbox{$N$ data points)}.}
\label{chi2}
\end{center}
\end{figure}
%%%%%%%%%%%%%%%%Figure 2 ("as")
\begin{figure}[f]
\begin{center}
\includegraphics[width=\textwidth]{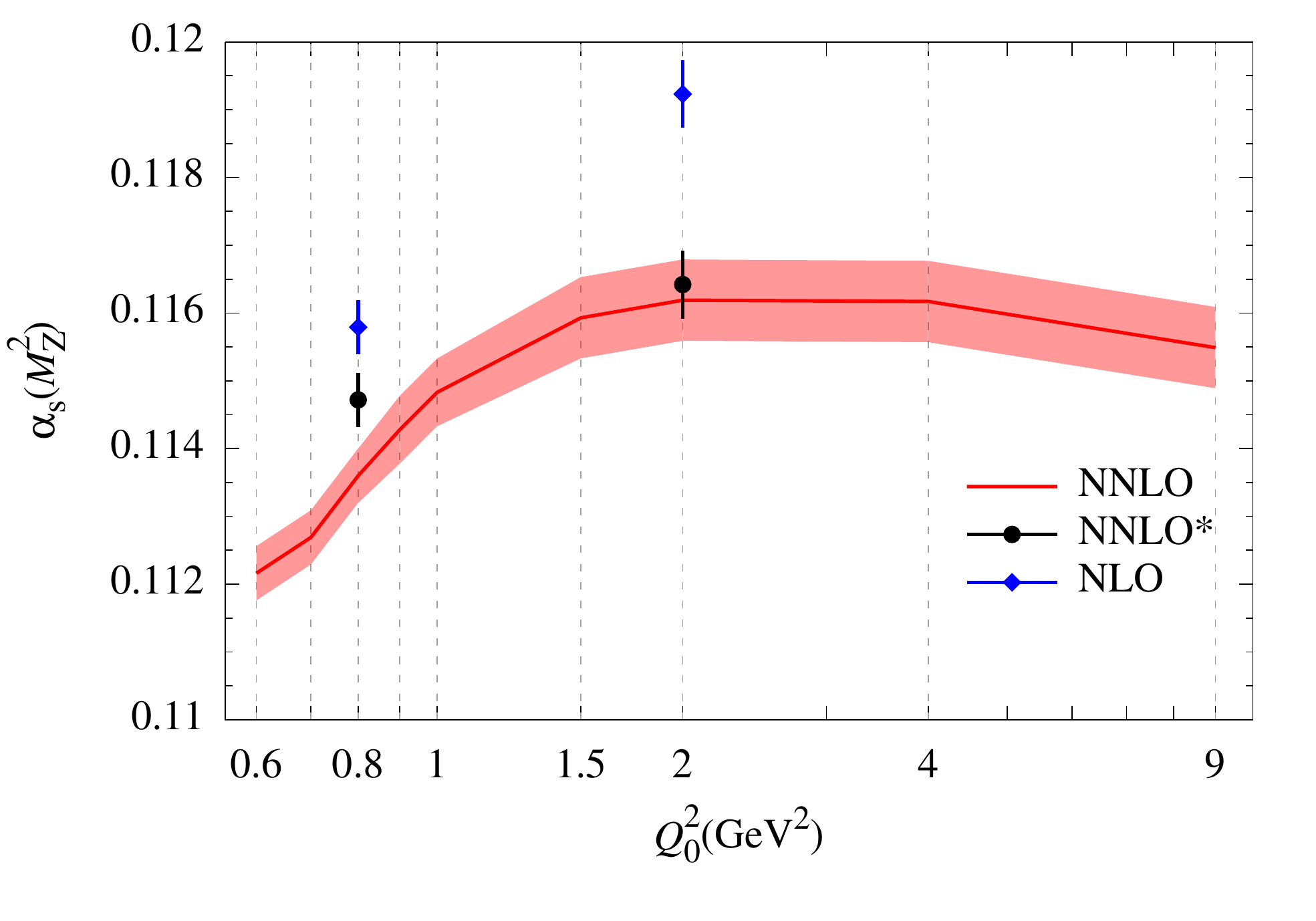}
\caption{The dependence of $\alpha_s(M_Z^2)$ on the variations of the input scale $Q_0$ as obtained in our NNLO analyses,
together with the $1\sigma$ uncertainty band ($\Delta\chi^2=1$). For illustration we show the sensitivity of two of our results when including jet and
charm data at NNLO as well, denoted by NNLO*, where NLO matrix elements have been (inconsistently) combined with NNLO
PDFs.  The NLO results at the input scales 0.8 GeV$^2$ (dynamical) and 2 GeV$^2$ (standard) are depicted as well.}
\label{as}
\end{center}
\end{figure}
%%%%%%%%%%%%%%%%%%Figure 3  ("scan")
\begin{figure}
\begin{center}
\includegraphics[width=\textwidth]{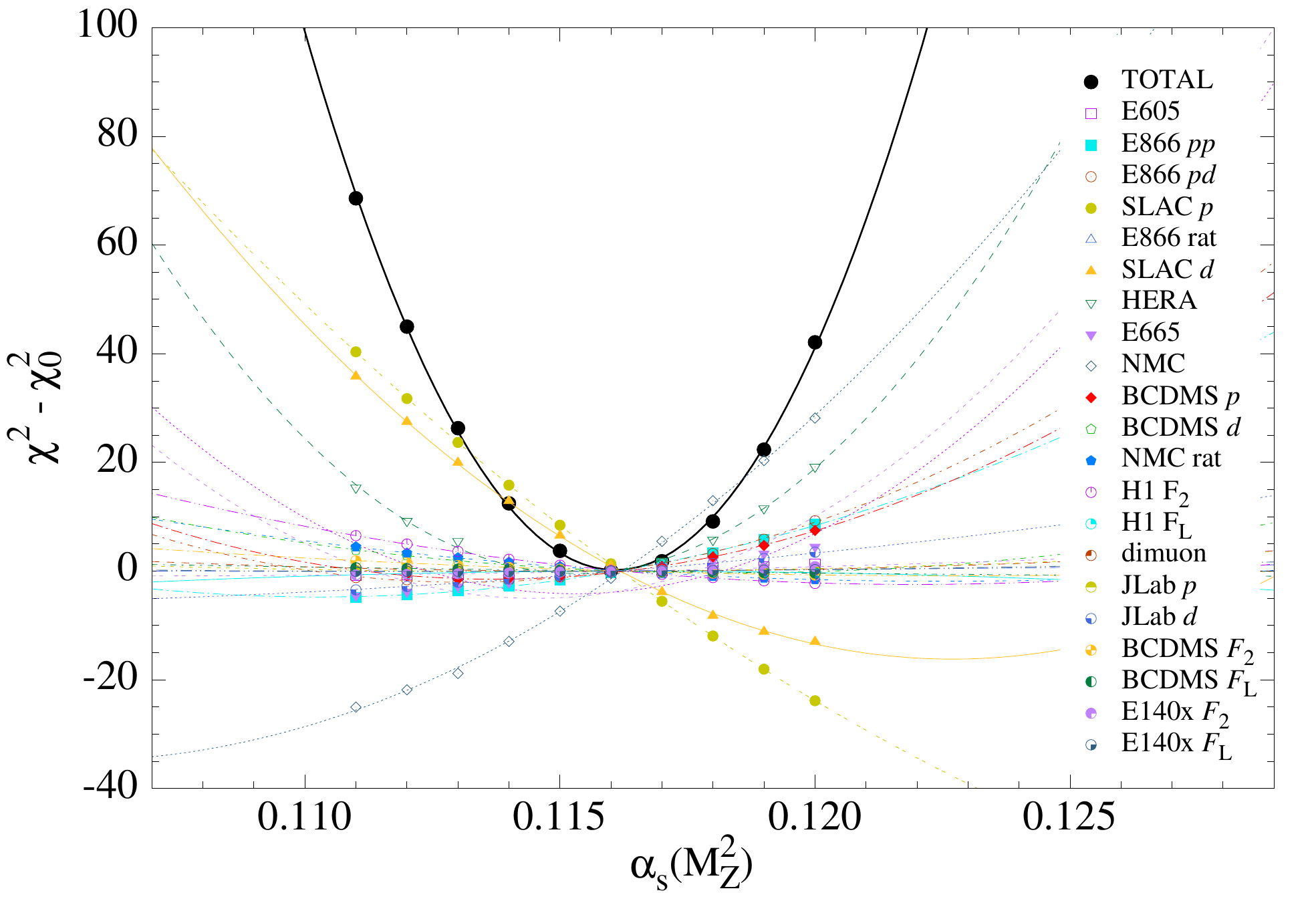}
\caption{The dependence of $\chi^2$ on $\alpha_s(M_Z^2)$ around the minimum $\chi_0^2$ for our standard NNLO analysis.}
\label{asscan_dyn}
\end{center}
\end{figure}
%%%%%%%%%%%%%%%%%%Figure 4  ("non-singlet")
\begin{figure}[f]
\begin{center}
\includegraphics[width=0.85\textwidth]{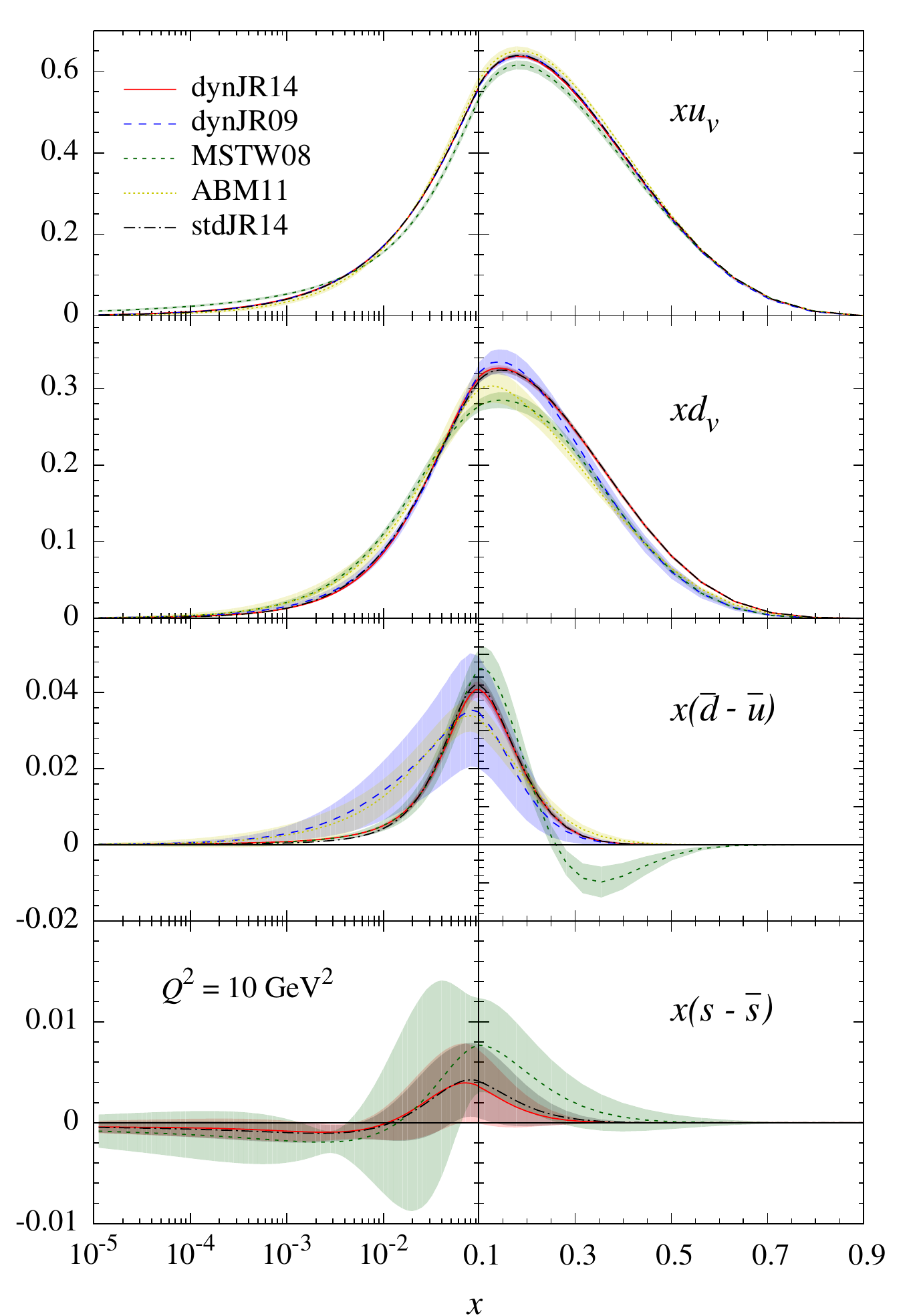}
\caption{Non-singlet NNLO distributions at $Q^2 = 10 \; {\rm GeV^2}$. Our present dynamical and standard PDFs are denoted by dynJR14 and stdJR14, respectively, and our previous dynamical JR09 results are taken from \cite{ref2}. These results are compared with the ones of ABM11 \cite{ref5} and the 3-flavor MSTW08 \cite{ref56} distributions.}
\label{nonsinglet}
\end{center}
\end{figure}
%%%%%%%%%%%%%%%%Figure 5  ("singlet")
\begin{figure}[f]
\begin{center}
\includegraphics[width=0.85\textwidth]{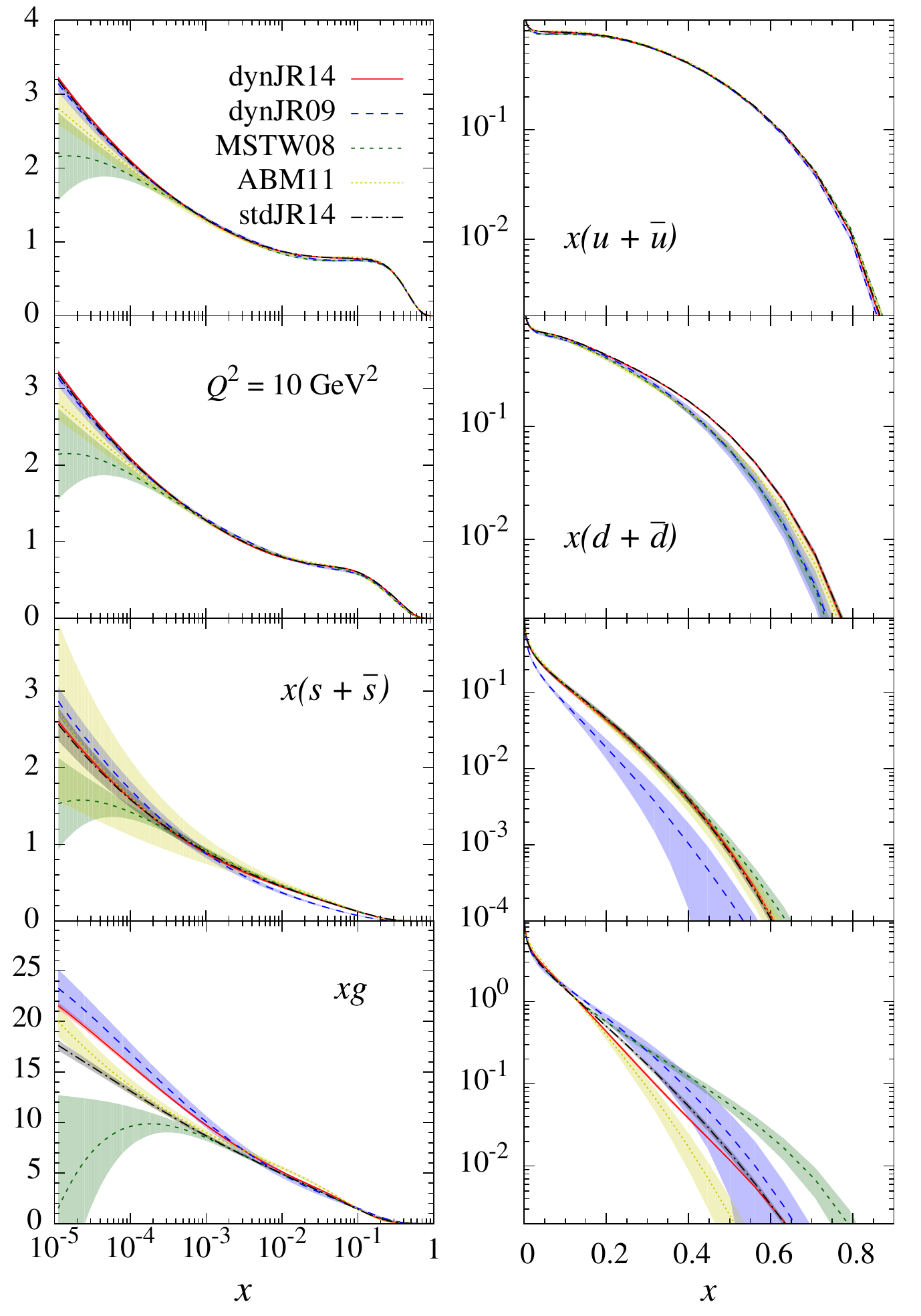}
\caption{As in Fig.~4 %\ref{nonsinglet} 
but for the singlet sector.}
\label{singlet}
\end{center}
\end{figure}
%%%%%%%%%%%%%%%Figure 6 ("ratios" = uvrat + dvrat + sigmarat + gluonrat)
\begin{figure}[bt]
\begin{center}
\includegraphics[width=0.495\textwidth]{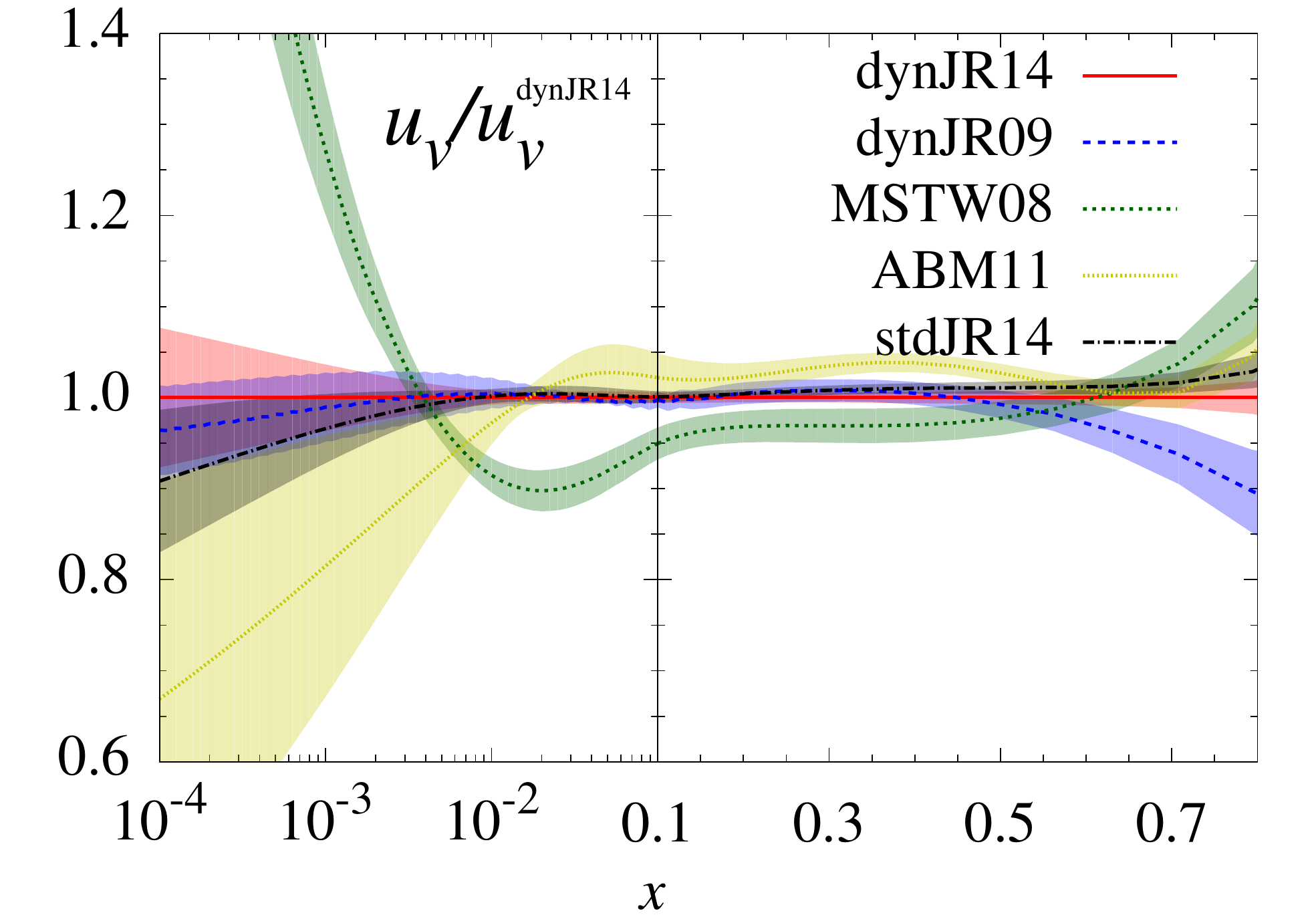}
\includegraphics[width=0.495\textwidth]{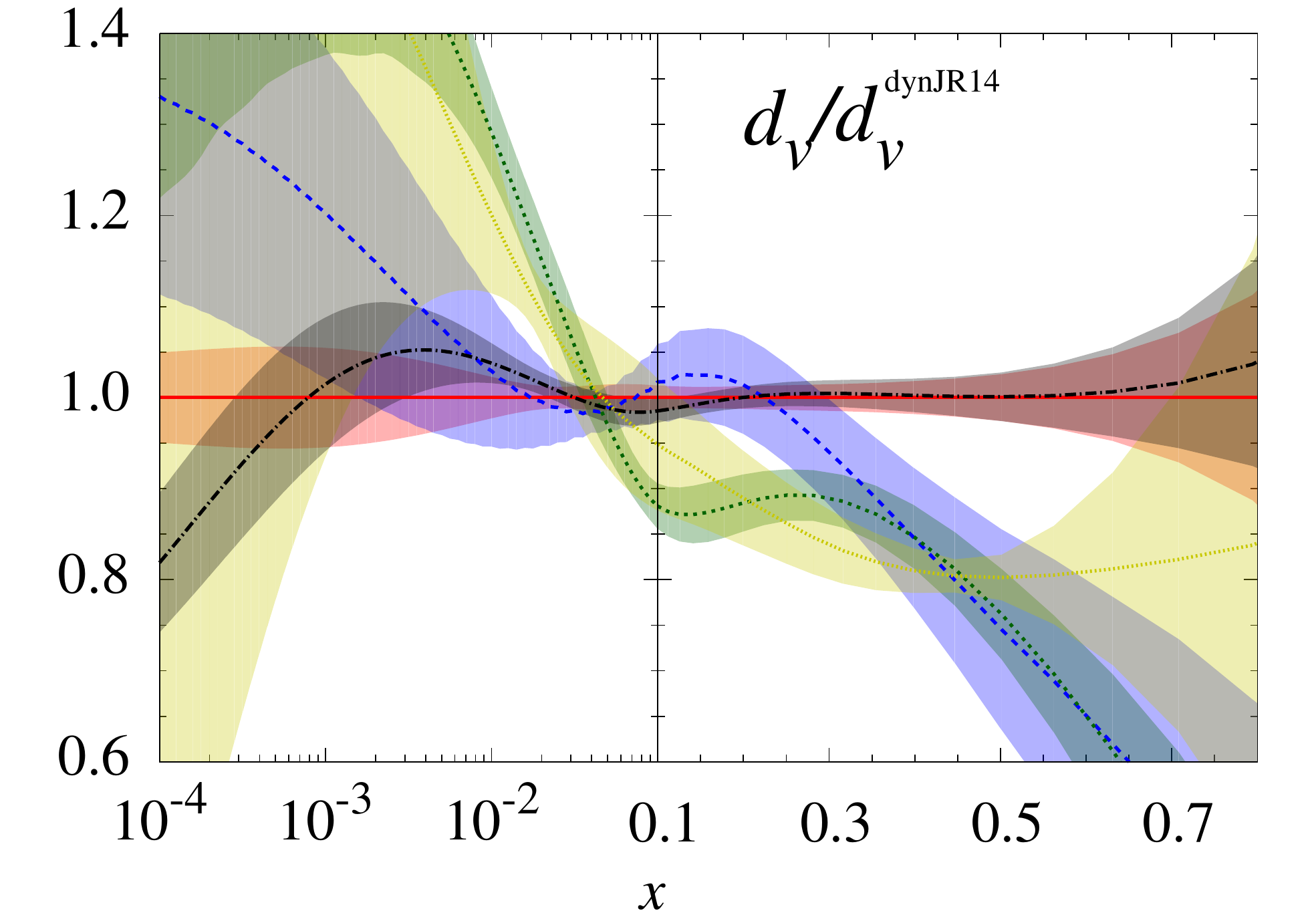}
\includegraphics[width=0.495\textwidth]{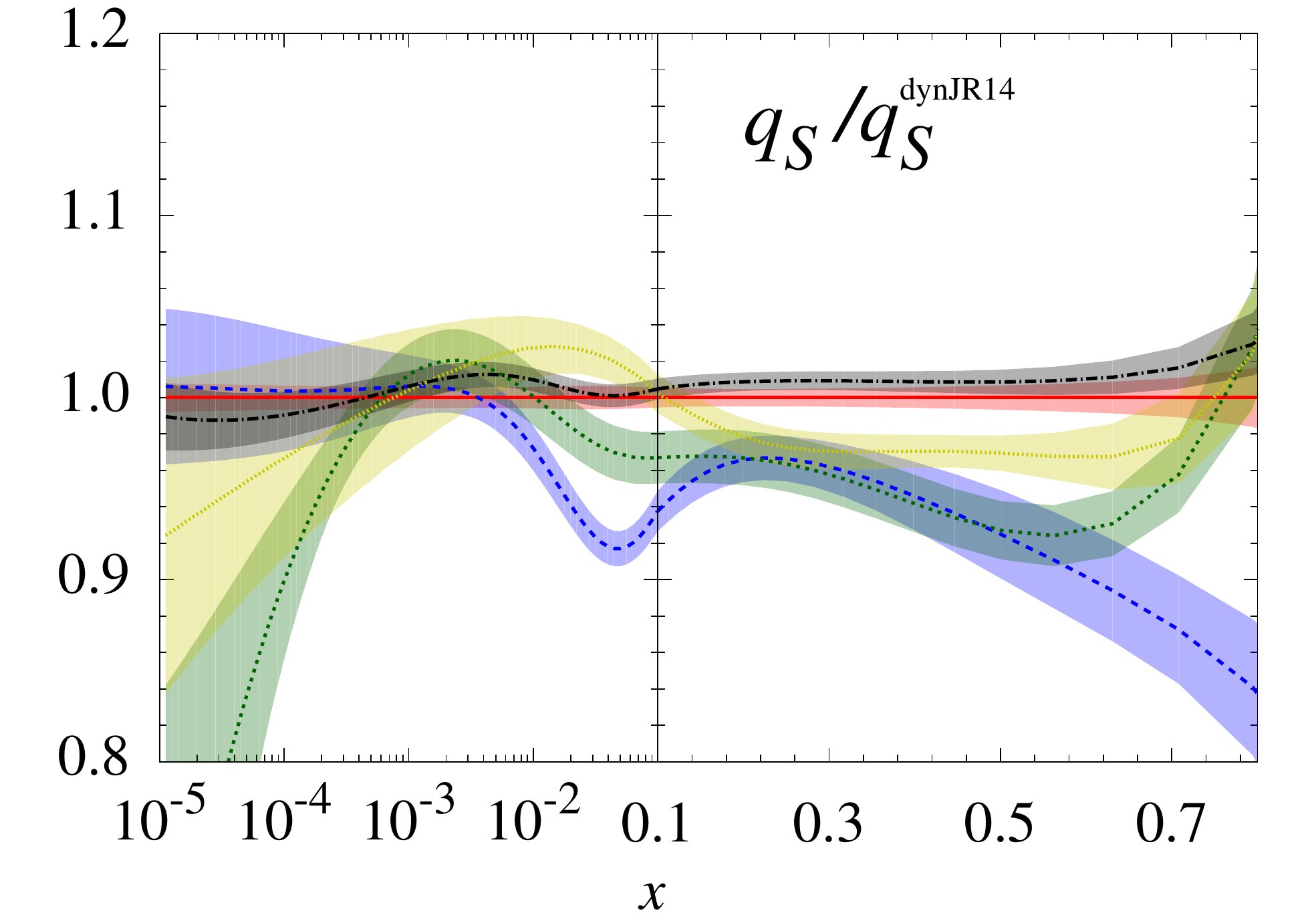}
\includegraphics[width=0.495\textwidth]{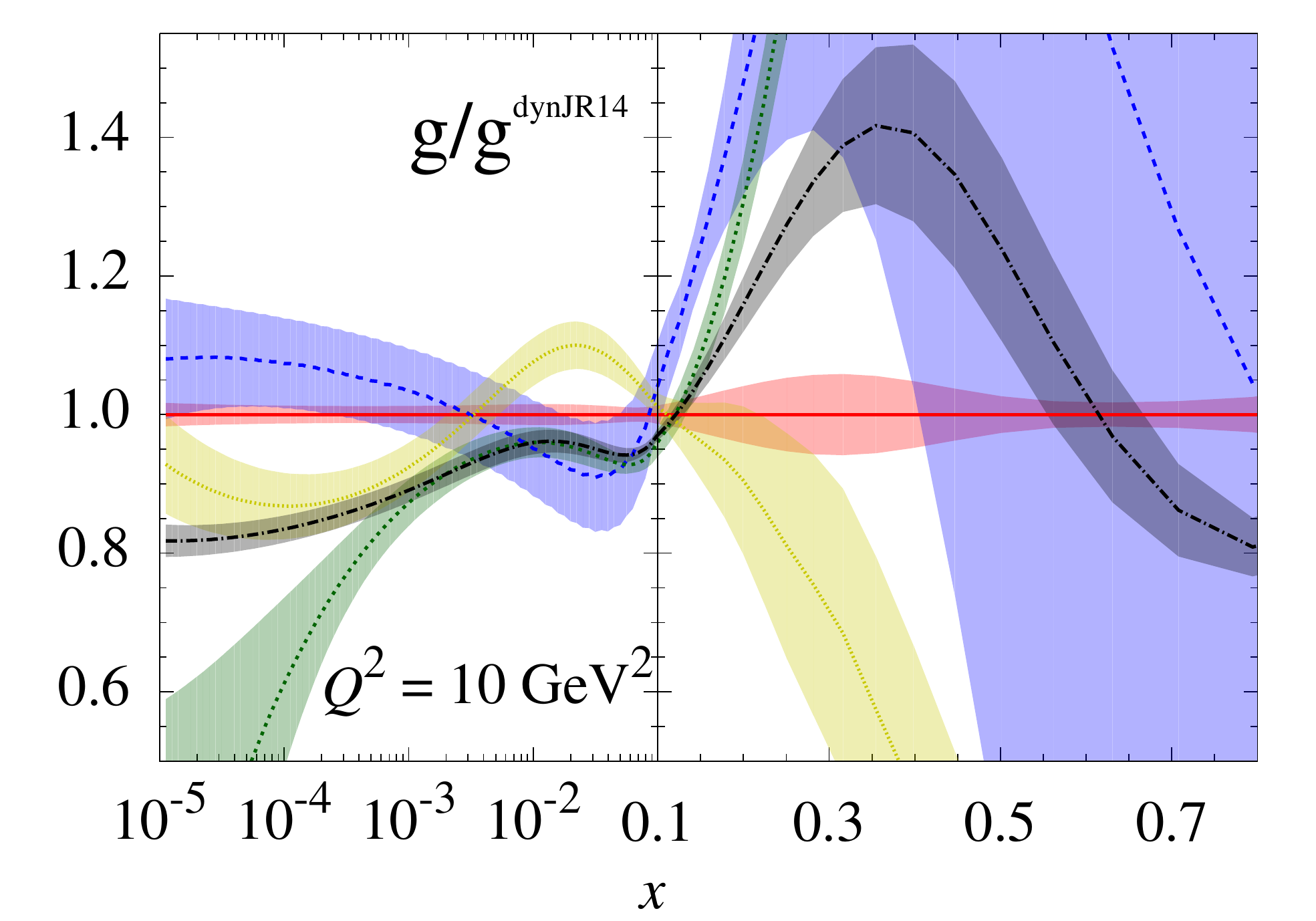}
\caption{As in Fig.~4 but for the ratios of NNLO PDFs with respect to our present dynamical result. The fermionic singlet distribution is defined by $q_S = u + \bar{u} + d + \bar{d} + s + \bar{s}$.}
\label{ratios}
\end{center}
\end{figure}
%%%%%%%%%%%%%%%Figure 7  ("NLOrat")
\begin{figure}[f]
\begin{center}
\includegraphics[width=0.93\textwidth]{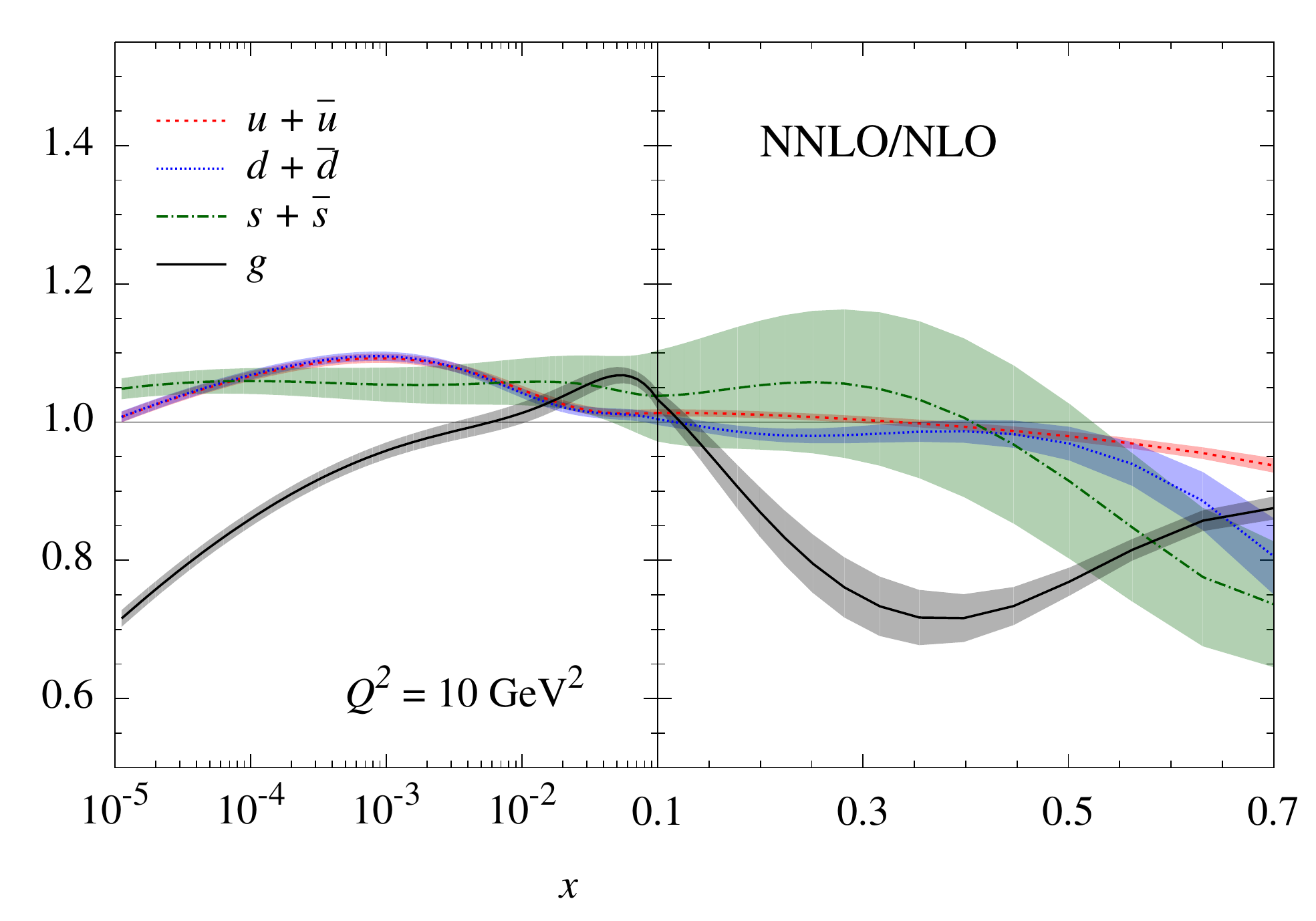}
\caption{The NNLO/NLO ratios of our dynamical JR14 parton distributions.}
\label{NLOrat}
\end{center}
\end{figure}
%%%%%%%%%%%%%%Figure 8  ("T4DIS")
\begin{figure}
\begin{center}
\includegraphics[width=0.85\textwidth]{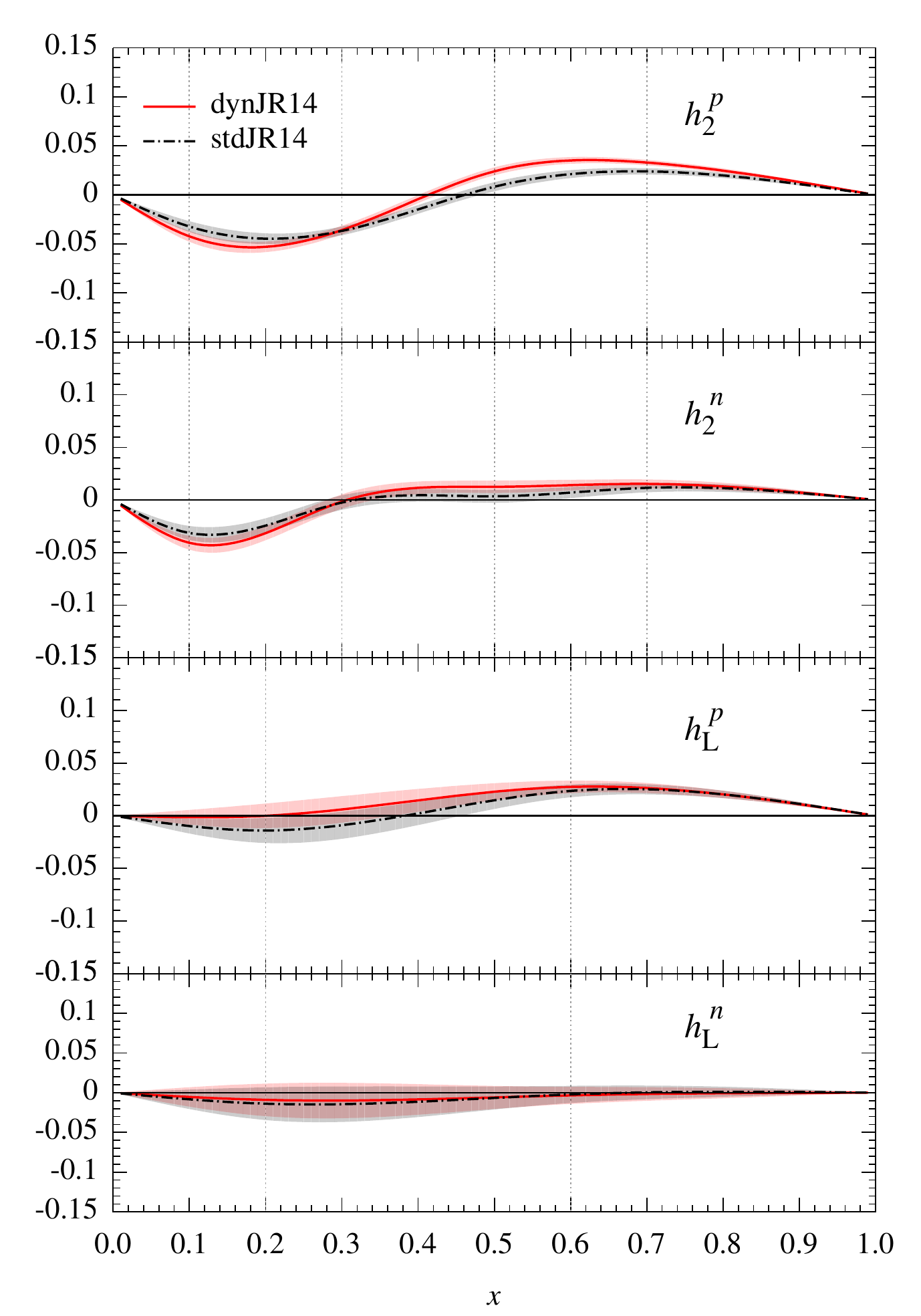}
\caption{The central values and $1\sigma$ bands of the higher-twist ($\tau=4$) contributions (in units of GeV$^2$) to the structure functions in Eq.(3) at NNLO.}
\label{T4DIS}
\end{center}
\end{figure}
%%%%%%%%%%%%%%Figure 9  ("FL")
\begin{figure}
\begin{center}
\includegraphics[width=\textwidth]{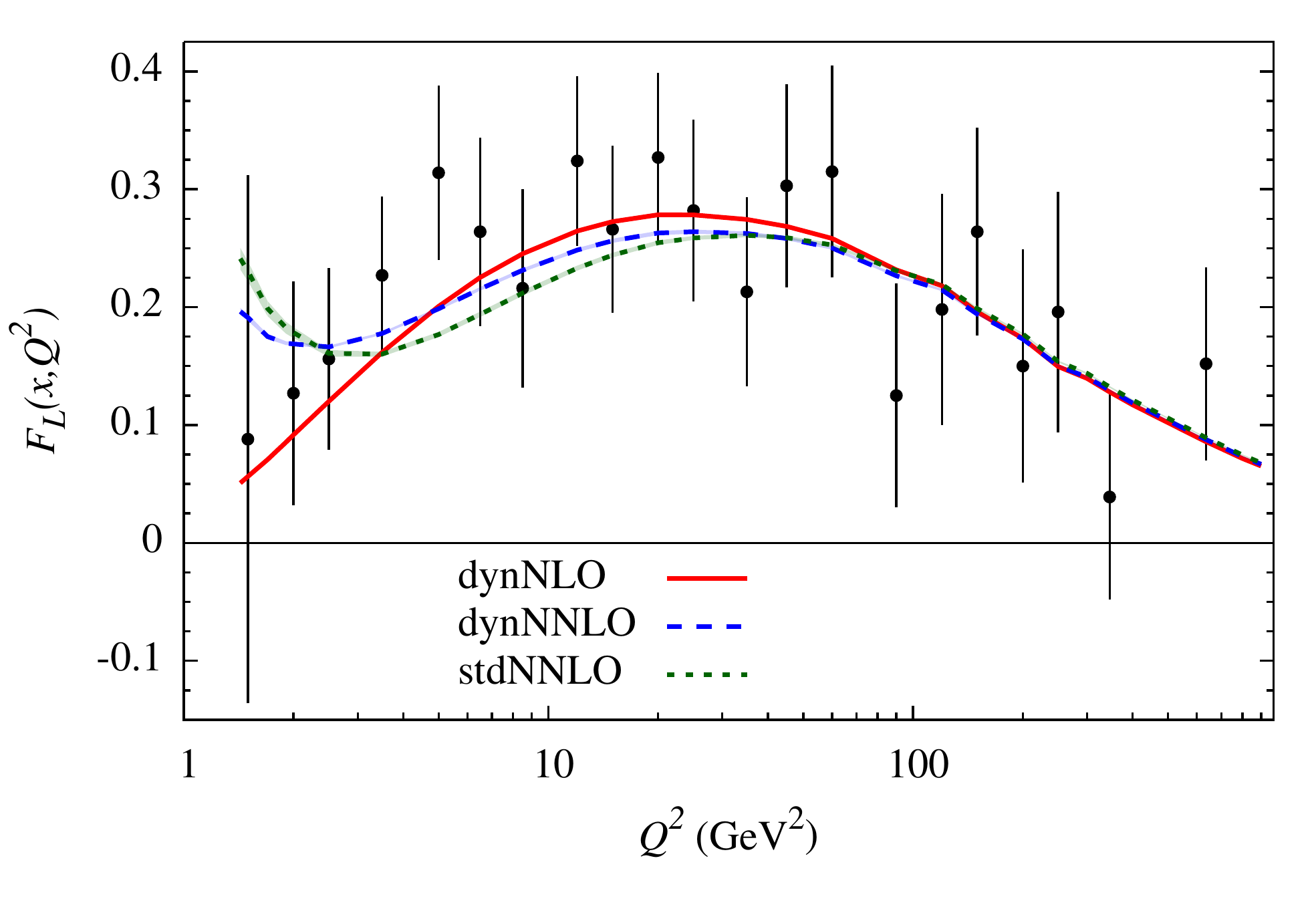}
\caption{The dynamical and standard predictions for the longitudinal structure function $F_L$.  The (small) $1\sigma$
uncertainty bands amount to only $0.5-1$\%.  The data \cite{ref41} cover a huge small-$x$ domain between 
$3 \times 10^{-5}$ (small  $Q^2$) and $2\times 10^{-2}$ (large $Q^2$).}
\label{FL}
\end{center}
\end{figure}

%%%%%%%%%%%%%%Figure 10 ("97 HERA")
\begin{figure}
\begin{center}
\includegraphics[width=\textwidth]{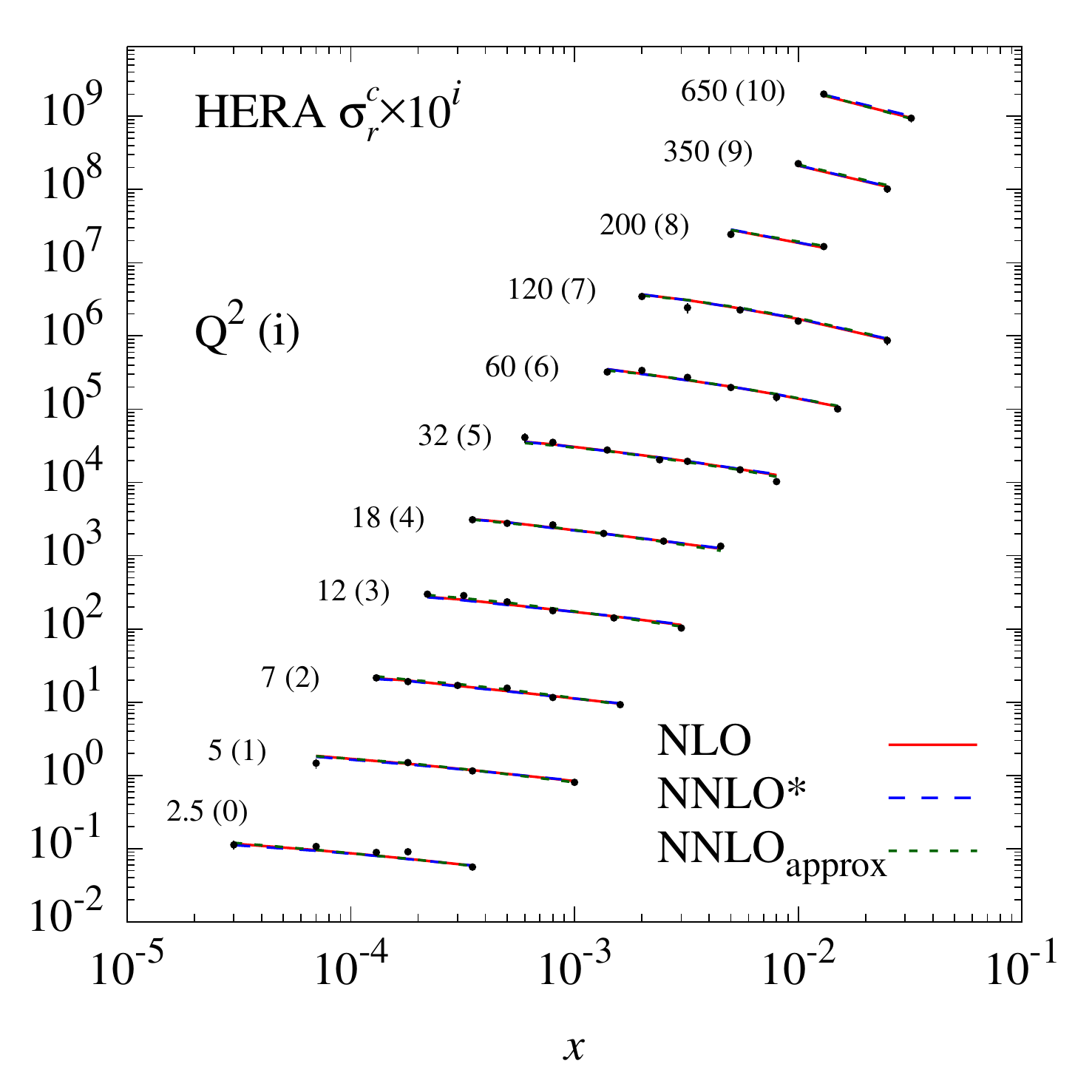}
\caption{Theoretical NLO predictions and NNLO expectations for DIS charm production at fixed values of $Q^2$ (GeV$^2$ units).
NNLO* refers to NNLO fits using (incorrectly) massive NLO sub-cross sections.  The optimal approximate NNLO$_{\rm approx}$
expectations are described in the text.  The shifts $r_j$ induced by the systematic errors (cf.~Eq.(5)) are included in the
theoretical predictions, and the errors shown are only statistical.  The HERA `reduced' cross-section data are taken 
from \cite{ref40}.}
\label{97HERA}
\end{center}
\end{figure}

%%%%%%%%%%%%%%%% Figure 11 ("jets")
\begin{figure}
\begin{center}
\includegraphics[width=0.49\textwidth]{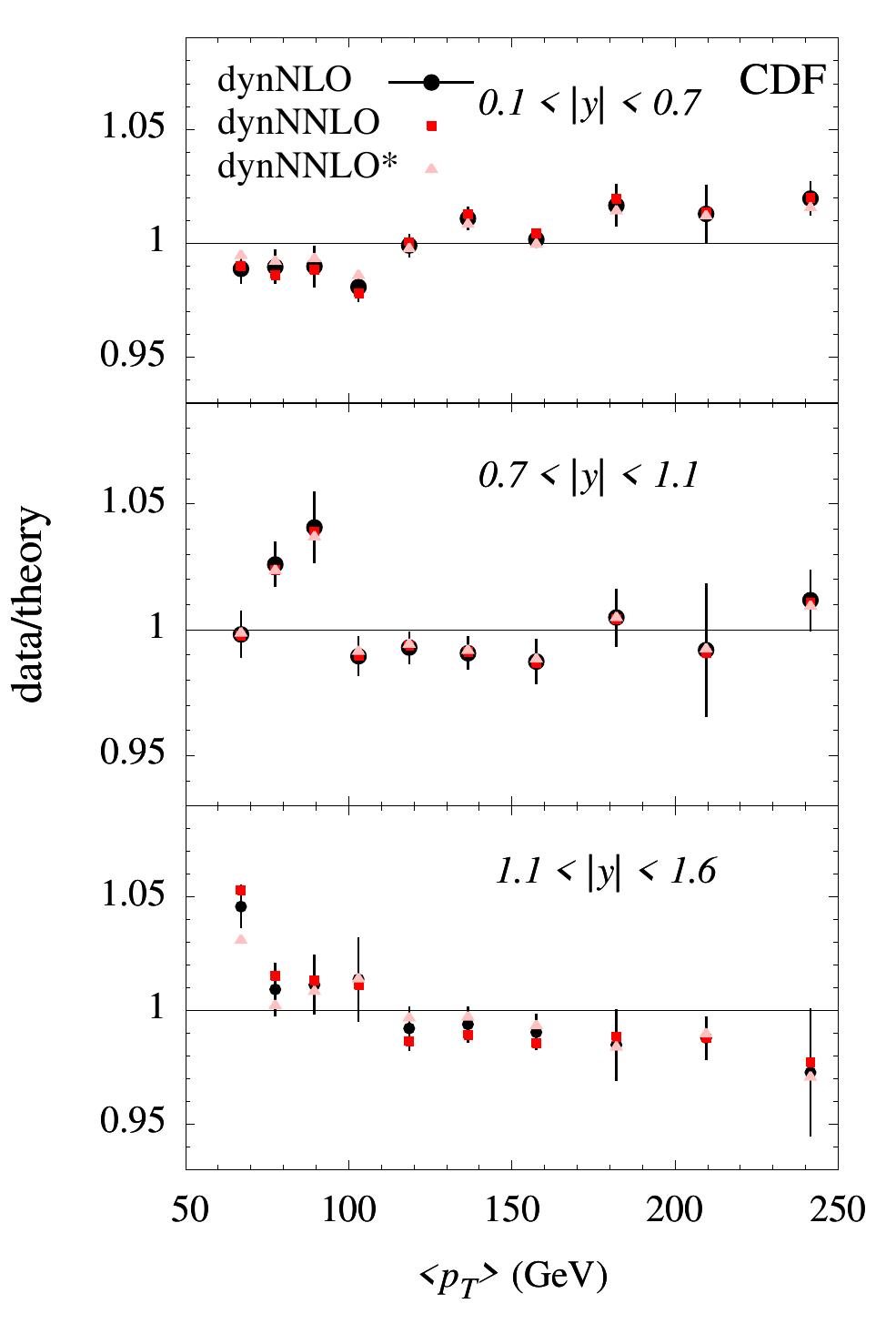}
\includegraphics[width=0.49\textwidth]{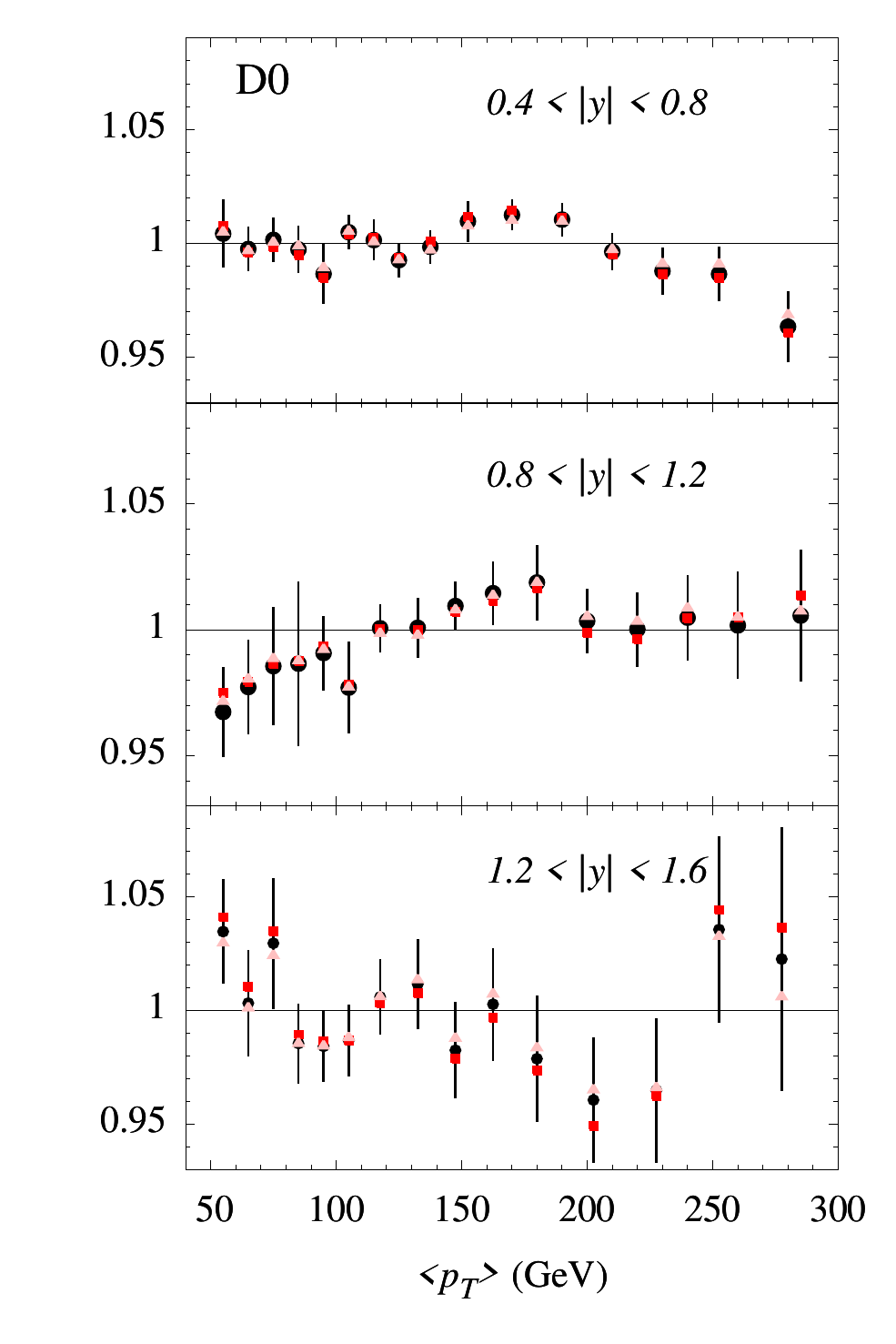}
\caption{Comparison of TeVatron hadronic jet-data \cite{ref48,ref49} with consistent theoretical dynamical NLO results,
and (inconsistent) NNLO expectations: NNLO* refers to fitting NNLO PDFs using NLO matrix elements, and NNLO uses
our optimal fixed NNLO PDFs (no fit) together with NLO matrix elements.}
\label{jets}
\end{center}
\end{figure}
%%%%%%%%%%%%%%Figure 12 ("ellipses" (=WZ + WPM)
\begin{figure}
\begin{center}
\includegraphics[width=0.7\textwidth]{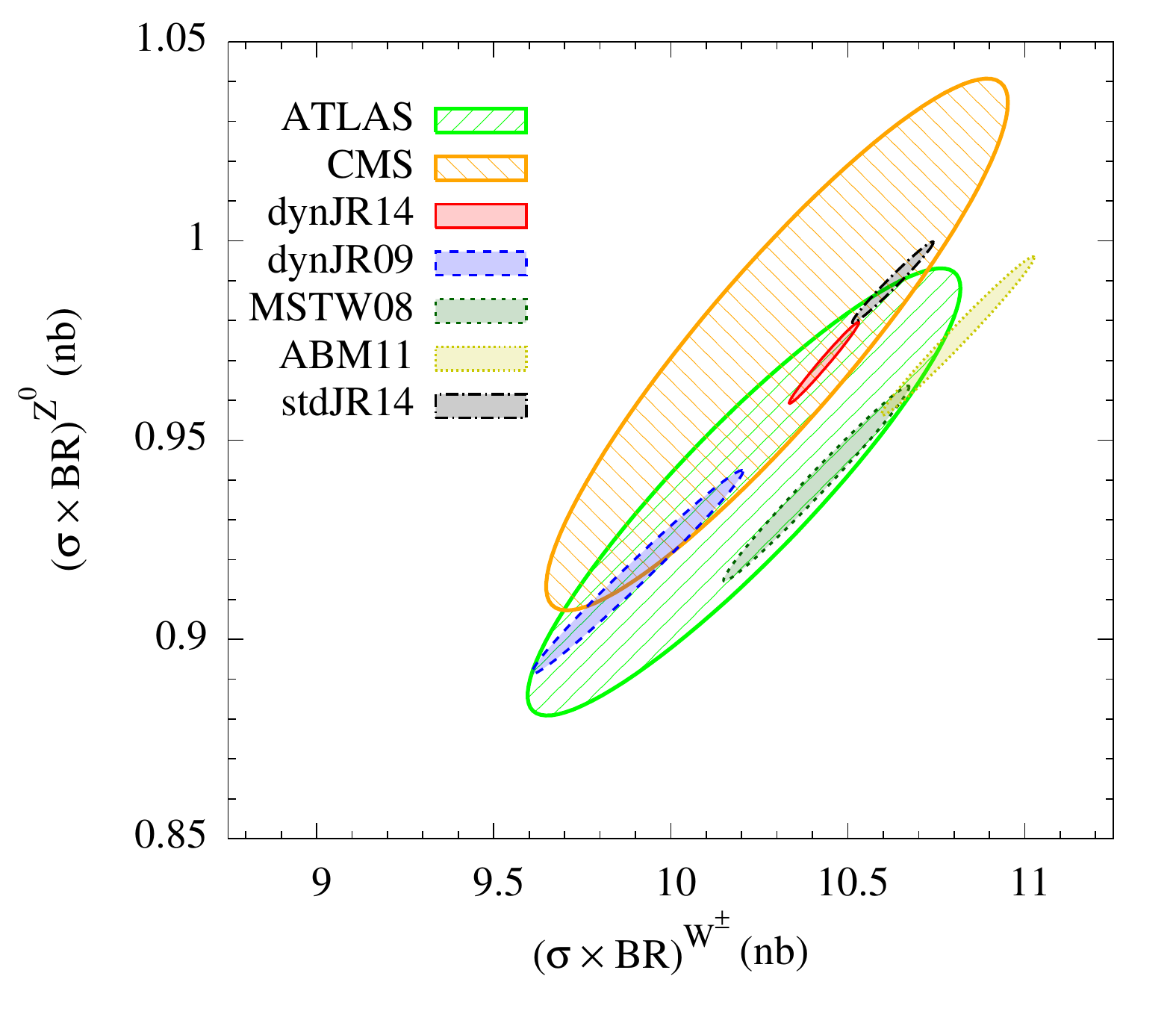}
\includegraphics[width=0.7\textwidth]{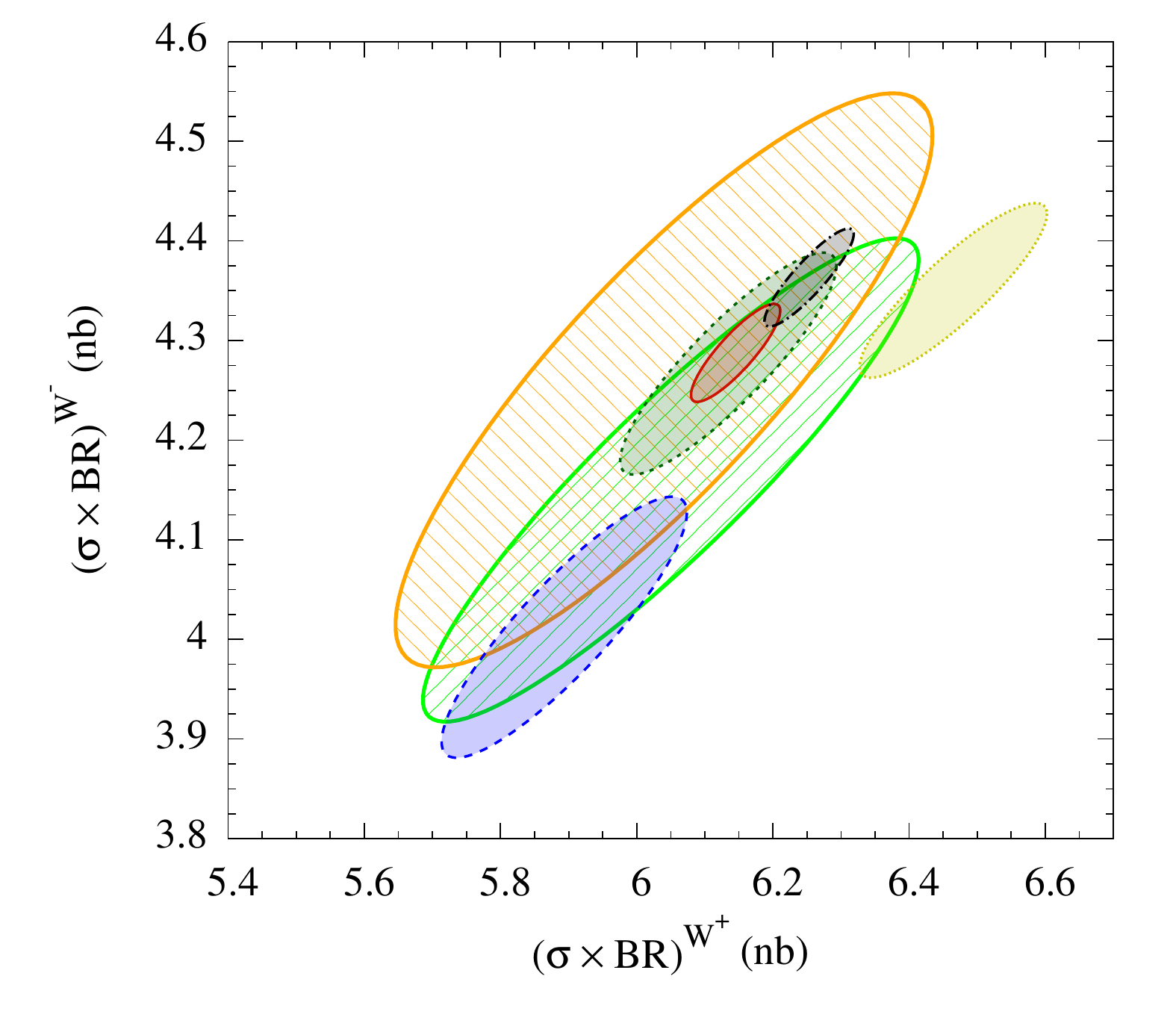}
\caption{1$\sigma$ ($\Delta \chi^2 = 2.3$) ellipses for $W^{\pm}\equiv W^+ +W^-$ vs. $Z^0$ and $W^{+}$ vs. $W^-$ total $pp$ cross sections at NNLO, compared to the LHC data from ATLAS \cite{ref67} and CMS\cite{ref68} at $\sqrt{s}=7$ TeV. For comparison we display the dynamical predictions of JR09 \cite{ref3} as well as of ABM11 \cite{ref5} and MSTW08 \cite{ref1}.}
\label{ellipses}
\end{center}
\end{figure}

%%%%%%%%%%%%%%%%%%%%%%%%%%%%%%%%%%%%%%%%%%%%%%%%%%%%%%%%%%%%%%%%
%%%%%%%%%%%%%%%% REFERENCES %%%%%%%%%%%%%%%%%%%%%%%%%%%%%%%%%%%%

\end{document}